\documentclass[pra,nofootinbib,twocolumn,superscriptaddress,longbibliography]{revtex4-2}
\usepackage{float,graphicx,multirow, amsmath,color,dsfont}
\usepackage{mathtools}
\usepackage{amsmath,bm}
\usepackage{amssymb,mathrsfs,dsfont}
\usepackage{amssymb,mathbbol}
\usepackage{amsfonts}
\usepackage{mathrsfs}
\usepackage{float}
\usepackage{xcolor,hyperref}
\definecolor{darkblue}{rgb}{0,0.0.1,0.3}
\definecolor{darkred}{rgb}{0.6,0.1,0}
\hypersetup{colorlinks,breaklinks,
linkcolor=darkred,urlcolor=darkblue,
anchorcolor=darkred,citecolor=
darkred,pdfauthor=ChAr, pdftitle=ChAr.Catalysis}
\usepackage{tikz}
\usepackage{mathbbol}
\usepackage{float}
\usepackage[normalem]{ulem}
\usepackage{ulem}

\newcommand{\ie}{\textit{i}.\textit{e}.}

\begin{document}
\title{Re-examination of the role of displacement and photon
catalysis operation in continuous variable measurement
device-independent quantum key distribution}
\author{Chandan Kumar}
\email{chandan.quantum@gmail.com}
\affiliation{Department of Physical Sciences,
Indian
Institute of Science Education and
Research Mohali, Sector 81 SAS Nagar,
Punjab 140306 India.}
\author{Arvind}
\email{arvind@iisermohali.ac.in}
\affiliation{Department of Physical Sciences,
Indian
Institute of Science Education and
Research Mohali, Sector 81 SAS Nagar,
Punjab 140306 India.}
\begin{abstract}
  We investigate the
benefits of  using $m$-photon catalysed two-mode
squeezed coherent ($m$-PCTMSC) state   in 
continuous variable measurement device-independent quantum
key distribution (CV-MDI-QKD).  To that end, we derive the
Wigner characteristic function of the $m$-PCTMSC state and show
that the 0-PCTMSC state is a Gaussian state and is an inferior
choice as compared to the zero photon catalyzed two-mode
squeezed vacuum state  for CV-MDI-QKD.  We carry out
the optimization of the secret key
rate with respect to all state parameters, namely
variance, transmissivity, and displacement.  Contrary to
many recent proposals, the results show that zero- and
single-photon catalysis operation provides
only a  marginal benefit in
improving the maximum transmission distance.  Secondly, 
we find  that
displacement offers no benefit  in improving 
CV-MDI-QKD. 
\end{abstract}
\maketitle
\section{Introduction}
Quantum key distribution (QKD) has the capability to create
a secure key shared  between two parties despite transmitting
data over an insecure quantum channel~\cite{gisin-rep-2002,
scarani-rmp-2009,Pirandola-2019}. This security  of QKD
protocols  is established  by the fundamental principles of
quantum mechanics.  Discrete variable(DV) as well
continuous variable(CV) quantum systems can be employed to
carry out QKD~\cite{Pirandola-2019}. Due to technical advantages,
 there has been significant increase in interest in   CV  QKD
  resulting in the rapid expansion of the field. Numerous CV-QKD protocols have
undergone theoretical investigation and practical validation
over the past two decades~\cite{Pirandola-2019}.

Inspired by the idea of entanglement swapping,
measurement-device-independent QKD (MDI-QKD) protocol was
proposed for discrete variable systems~\cite{Braunstein-prl-2012,Lo-prl-2012}.
 The MDI-QKD protocols  are   immune to attack on
detectors. The corresponding MDI-QKD for CV system was
 constructed and~\cite{Pirandola-np-2015, Li-pra-2014, Xiang-pra-2014}
several efforts have been made to enhance their efficiency
in terms of maximum possible transmission distance over
which secure QKD can be established.

Non-Gaussian operations have been shown to enhance the
performance of quantum
teleportation~\cite{tel2000,dellanno-2007,tel2009,catalysis15,catalysis17,wang2015,tele-2023,noisytele},
quantum metrology~\cite{gerryc-pra-2012,josab-2012,braun-pra-2014,josab-2016,pra-catalysis-2021,ill2008,ill2013,metro22,
metro-thermal-arxiv,ngsvs-arxiv} and quantum key
distribution~\cite{qkd-pra-2013,qkd-pra-2019,qk2019,zubairy-pra-2020}.
Non-Gaussian operations such as photon subtraction, photon
addition, and photon catalysis (PC) operations were
considered in the context of CV-MDI-QKD
operations~\cite{Ma-pra-2018,chandan-pra-2019,
catcvmdiqkd2020,catalysis2021}.

In this work, we consider the application of PC operation on
the family of two mode squeezed coherent (TMSC) states and
utilize them  as resource states in CV-MDI-QKD.  We first
analytically derive the Wigner characteristic function of
the $m$-photon catalyzed TMSC ($m$-PCTMSC) state, which is
used to evaluate the covariance matrix. As a special case,
we find that the Wigner characteristic function of the
0-PCTMSC state is Gaussian, while for higher values of $m$,
the state is non-Gaussian.  The Wigner characteristic
function allows us to compute the covariance matrix required
for the evaluation of the secret key rate for the protocol
with $m$-PCTMSC state as the resource state.  We optimized
the secret key rate with  respect to the  state parameters,
namely, variance, transmissivity, and displacement  for a
given value of $m$.

The results show that PC operations performed by Alice are
not particularly useful and yield to only a marginal
advantage.  Displacement coupled with PC operation is not a
useful operation as far as QKD is concerned, which is
similar to our recent result on displacement coupled to
photon subtraction~\cite{long-qkd}.  In fact, the optimal
variance for large distances is on the low range where both
PC and displacement are not useful. These results are in
contrast with several earlier pieces of published
work~\cite{catcvmdiqkd2020,catalysis2021}, where the
advantage obtained  was an artifact of working at high
variance.

We were able to reach these
conclusions because of our global approach and the fact that
we had an analytical expression for the Wigner characteristic
function for the $m$-PCTMSC state.
Our results will directly impact other similar studies on
photon catalysis such as virtual zero photon
catalysis~\cite{virtualpc} and underwater
CV-MDI-QKD~\cite{waterpc}.

The rest of the paper is organized in the following manner.
In Sec.~\ref{wig:pctmsc}, we commence by deriving the Wigner
function of $m$-PCTMSC state. Section~\ref{cvmdi} introduces
the $m$-PCTMSC state   CV-MDI-QKD protocol.  In
Sec.~\ref{numerical}, we present a detailed numerical
simulation and optimize the secret key rate over the state
parameters.   Finally, Sec.~\ref{sec:conc} provides
concluding remarks and future research directions.
\section{Wigner characteristic function and covariance
matrix of the $m$-PCTMSC state}\label{wig:pctmsc}
\begin{figure}[h!] 
\centering
\includegraphics[scale=1]{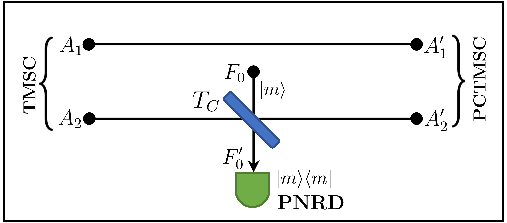}
\caption{  Schematic illustrating the generation of the
PCTMSC state from the TMSC state. A beam splitter with
transmissivity $T_C$ is used to interfere  the ancillary
mode $F_0$ in Fock state $|m\rangle$ and the mode $A_2$ of
the TMSC state. An $m$-PCTMSC state is successfully
generated when $m$ photons are detected by a photon number
resolving detector. } \label{pstmsc_sub}
\end{figure}
The Wigner characteristic function corresponding to an
$n$-mode quantum system state, as represented by the density
operator $\hat{\rho}$, can be expressed as follows:

\begin{equation}\label{wigdef}
\chi(\Lambda) = \text{Tr}[\hat{\rho} \, \exp(-i \Lambda^T
\Omega \hat{\xi})],\quad
\Omega = {\displaystyle \bigoplus_{k=1}^{n}}\begin{pmatrix}
0&1\\
-1&0.
\end{pmatrix}
\end{equation}
where  $\hat{\xi} = (\hat{q_1}, \hat{p_1},\dots, \hat{q_n},
\hat{p_n})^T$, and  $\Lambda = (\Lambda_1, \Lambda_2, \dots,
\Lambda_n)^T$ with  $\Lambda_k = (\tau_k, \sigma_k)^T \in
\mathcal{R}^2$.

States with Gaussian Wigner characteristic function are
known as Gaussian states. Such states can be uniquely
specified via the first order moments and second order
moments. First order moments is defined as $  \bm{d}=
\langle \hat{\xi} \rangle $. The second order moments can be
written in the form of a  matrix  called  covariance matrix
$\Sigma$, whose entries are given by
\begin{equation}
	(\Sigma)_{ij} = \frac{1}{2}\langle\{
	\hat{\xi_i},\hat{\xi_j} \} \rangle - \langle
	\hat{\xi_i}\rangle \langle \hat{\xi_j}\rangle .
\end{equation}

Figure~\ref{pstmsc_sub}  shows the schematic for generating
the PCTMSC state. We suppose that the modes $A_1$ and $A_2$
are initialized to the TMSC state.   To generate TMSC state,
we consider two uncorrelated modes, each of them initialized
to a coherent state. Such a state is Gaussian and,
therefore, can be described by the following displacement
vector and covariance matrix (in shot noise units):
\begin{equation}
\bm{d} = ( d,0,d,0)^T,\quad \Sigma= \mathbb{1}_4,
\label{disp_cov}
\end{equation}
where $\mathbb{1}_4$ is the $4 \times 4$ identity matrix.   
To generate a TMSC state, these modes are passed via a
two-mode nonlinear optical down converter, whose effect can
be modeled by the following symplectic transformation
corresponding to two mode squeezing operation:
\begin{equation}\label{eq:tms}
S_{A_1 A_2}(\lambda) = \frac{1}{\sqrt{1-\lambda ^2}} \begin{pmatrix}
\,\mathbb{1}_2& \lambda \,\mathbb{Z} \\
\lambda \,\mathbb{Z}&  \,\mathbb{1}_2
\end{pmatrix},
\end{equation}
where $\mathbb{Z} = \text{diag}(1,\, -1)$ and $\lambda =
\tanh (r)$ with $r$ being the squeezing parameter.
The displacement vector  of the TMSC state can be evaluated as
\begin{equation}\label{tmscmean}
\bm{d'} = S_{A_1 A_2 }(\lambda) \bm{d}  =
\frac{d(1+\lambda)}{\sqrt{1-\lambda ^2}} \begin{pmatrix}
1  \\
0 \\
1  \\
0 \\
\end{pmatrix}.
\end{equation}
Similarly, the covariance matrix of the TMSC state can be calculated as
\begin{equation}\label{tmsccov}
\begin{aligned}
\Sigma' & =S_{A_1 A_2 }(\lambda)  \Sigma S_{A_1 A_2 }(\lambda)^T,\\
&= \frac{1}{ 1-\lambda ^2} \begin{pmatrix}
(1+\lambda ^2)\,\mathbb{1}_2& 2\lambda \,\mathbb{Z} \\
2\lambda \,\mathbb{Z}&  (1+\lambda ^2)\,\mathbb{1}_2
\end{pmatrix}.
\end{aligned}
\end{equation}
For the TMSC state, which is Gaussian and specified by the
displacement vector~(\ref{tmscmean}) and covariance
matrix~(\ref{tmsccov}), Eq.~(\ref{wigdef}) acquires a simple
form~\cite{Weedbrook-rmp-2012, olivares-2012}:
\begin{equation}\label{wigc}
\chi(\Lambda) =\exp[-\frac{1}{2}\Lambda^T (\Omega \Sigma'
\Omega^T) \Lambda- i (\Omega \bm{d'} )^T\Lambda].
\end{equation}
Using the above equation, the Wigner characteristic function
of the TMSC state, $\chi_{A_1 A_2}(\Lambda_1, \Lambda_2)$,
can be easily calculated.
We  consider an ancilla mode $F_0$  initialized to the Fock
state $|m\rangle$. To interfere the mode $F_0$ with  the
mode $A_2$ of the TMSC state,  a beam splitter with
transmissivity $T_C$ is utilized. Before the interference,
the Wigner characteristic function of the three mode system
is given by
\begin{equation}
\chi_{A_1 A_2F_0}(\Lambda ) = \chi_{A_1 A_2}(\Lambda_1,
\Lambda_2)\chi_{|m\rangle}(\Lambda_3), 
\end{equation} 
where   $\chi_{|m\rangle}(\Lambda_3)$ is the Wigner
characteristic function of the Fock state:
\begin{equation}\label{charfock}
\chi_{|m\rangle}(\Lambda_3)=\exp  \left[-
\frac{\tau_3^2}{2}-\frac{\sigma_3^2}{2}
\right]\,\mathcal{L}_{m}\left(  \tau_3^2 + \sigma_3^2
\right),
\end{equation}
where $\mathcal{L}_m(x)$ is the Laguerre polynomial.
The three modes get entangled by the beam splitter's action,
and    the modified Wigner characteristic function is given
by
\begin{equation}
\chi_{A'_1 A'_2F'_0}(\Lambda ) = \chi_{A_1
A_2F_0}\left([\mathbb{1}\oplus B_{A_2 F_0}(T_C)]^{-1}\Lambda
\right),
\end{equation}
where $B_{A_2 F_0}(T_C)$ represents the beam splitter
transformation  given by
\begin{equation}\label{beamsplitter}
B_{A_2 F_0}(T_C) = \begin{pmatrix}
\sqrt{T_C} \,\mathbb{1}& \sqrt{1-T_C} \,\mathbb{1} \\
-\sqrt{1-T_C} \,\mathbb{1}& \sqrt{T_C} \,\mathbb{1}
\end{pmatrix}.
\end{equation}

A photon number resolving detector  is employed to measure
the mode $F'_0$. The measurement is described by the
positive operator valued measure $\{ \Pi_m = |m\rangle
\langle m|, \mathbb{1}-\Pi_m \}$.    The detection of $m$
photons indicates a successful $m$-photon catalysis
operation. The unnormalized Wigner characteristic function
of the  $m$-PCTMSC  state is  
\begin{equation}\label{wigred}
\begin{aligned}
\widetilde{ \chi}^{(m)}_{A_1 A'_2} (\Lambda_1, \Lambda_2)
=\frac{1}{  \pi} \int d^2 \Lambda_3 \, \chi_{A'_1
A'_2F'_0}(\Lambda )    
\chi_{|m\rangle }(-\Lambda_3) .
\end{aligned}
\end{equation}
\subsection{Zero photon catalyzed TMSC state}\label{wig:zpc}
We carry out a  direct
calculation of the integral~(\ref{wigred}) for $m=0$
to evaluate  the Wigner characteristic function  for 0-PCTMSC state:
 \begin{equation}\label{unzpc}
\begin{aligned}
\widetilde{ \chi}^{(0)}_{A_1 A'_2} (\Lambda_1, \Lambda_2) &
= \frac{\left(1-\lambda ^2 \right)  }{1-\lambda ^2 T_C } 
\exp \left[ -d^2  \frac{(\lambda +1)^2 \left(1-T_C\right)}{4
\left(1-\lambda ^2 T_C\right)}\right]\\
& \times \exp[-\frac{1}{2}\Lambda^T (\Omega
\,\Sigma_{PC}^{(0)} \,\, \Omega^T) \Lambda- i (\Omega
\,\,\bm{d}_{PC}^{(0)} )^T\Lambda],
\end{aligned}
\end{equation}
where \begin{equation}\label{meanzpc}
\bm{d}_{PC}^{(0)} = \frac{d \sqrt{1-\lambda ^2}}{1-\lambda
^2 T_C} \begin{pmatrix}
(1+\lambda  T_C)  \\
0 \\
(1+\lambda )   \sqrt{T_C}  \\
0 \\
\end{pmatrix},
\end{equation}
and
\begin{equation}\label{covzpc}
\Sigma_{PC}^{(0)} = \frac{1}{ 1- \overline{\lambda}^2} \begin{pmatrix}
(1+\overline{\lambda}^2)\,\mathbb{1}_2& 2\overline{\lambda} \,\mathbb{Z} \\
2\overline{\lambda} \,\mathbb{Z}&  (1+\overline{\lambda}^2)\,\mathbb{1}_2
\end{pmatrix}, \quad      \overline{\lambda}= \lambda \sqrt{T_C}.
\end{equation}

The success probability of  zero photon catalysis  can be
calculated from Eq.~(\ref{unzpc}) as
\begin{equation}
\begin{aligned}
P_{PC}^{(0)} =\widetilde{ \chi}^{(0)}_{A_1 A'_2}&(\Lambda_1, \Lambda_2)  
\bigg|_{\substack{\tau_1= \sigma_1=0\\ \tau_2= 
\sigma_2=0}} = \frac{\left(1-\lambda ^2 \right)  }{1-\lambda ^2 T_C } \\
&\times \exp \left[ -d^2  \frac{(\lambda +1)^2
\left(1-T_C\right)}{4 \left(1-\lambda ^2 T_C\right)}\right].
\end{aligned}
\end{equation}
Hence,  the normalized Wigner characteristic function for
0-PCTMSC can be expressed as follows:
\begin{equation}\label{normzpc}
\begin{aligned}
\chi^{(0)}_{A_1 A_2'}(\Lambda_1, \Lambda_2) &=\left(P_{PC}^{
(0) }\right)^{-1}\widetilde{ \chi}^{(0)}_{A_1
A'_2}(\Lambda_1, \Lambda_2), \\
& = \exp[-\frac{1}{2}\Lambda^T (\Omega \,\Sigma_{PC}^{(0)}
\,\, \Omega^T) \Lambda- i (\Omega \,\,\bm{d}_{PC}^{(0)}
)^T\Lambda].
\end{aligned}
\end{equation}
Since the Wigner characteristic function is Gaussian, the
0-PCTMSC state is a Gaussian state with mean
$\bm{d}_{PC}^{(0)}$~(\ref{meanzpc}) and covariance matrix
$\Sigma_{PC}^{(0)}$~(\ref{covzpc}).  It is important to note
that the covariance matrix of the 0-PCTMSC
state~(\ref{meanzpc}) is independent of the displacement
$d$.
By substituting $d=0$ in the above equation, we can
calculate the Wigner characteristic function for $0$-PCTMSV
states. The form of the covariance matrix~(\ref{covzpc})
suggests that the 0-PCTMSV state is another Gaussian state
with a small variance.  
Additionally, if we take the limit of $T_C$ approaching
unity in Eq.~(\ref{normzpc}), we can obtain the Wigner
characteristic function of the TMSC state.

The success probability for 0-PCTMSV can be obtained by
setting $d=0$ in the above equation:
\begin{equation}
P_{PC}^{(0)} =\frac{\left(1-\lambda ^2 \right)  }{1-\lambda ^2 T_C }.
\end{equation} Since 
\begin{equation}
0 \leq   \frac{(\lambda +1)^2 \left(1-T_C\right)}{4 \left(1-\lambda ^2 T_C\right)} \leq 1,
\end{equation}
the success probability of the 0-PCTMSC state is less than
the 0-PCTMSV state. 
 We would also like to remark that in
Ref.~\cite{catalysis2021}, it has been claimed that the
0-PCTMSC state is a non-Gaussian state, which our explicit
calculation shows is an incorrect statement.  
\subsection{Calculation for $m$-photon catalysis}
To evaluate the integral~(\ref{wigred}) for the general
case,  we need to express the Laguerre polynomial present in
the Wigner characteristic function of the Fock state $|m
\rangle$   as follows:
\begin{equation}
\mathcal{L}_{m}\left(\tau_3^2+\sigma_3^2\right)
=\frac{1}{m!} \frac{\partial^{2m}}{\partial u ^{m}\partial
v^{m}} e^ {uv+u  (\tau_3+i\sigma_3)-v (\tau_3-i\sigma_3) }
\bigg|_{\substack{u =0\\ v=0}}.
\end{equation}
When the aforementioned equation is substituted in
Eq.~(\ref{wigred}), a Gaussian integral is obtained that
evaluates to
\begin{equation}\label{reducedcal}
\begin{aligned}
\widetilde{ \chi}^{(m)}_{A_1 A'_2}& = P_{PC}^{(0)}
\chi^{(0)}_{A_1 A_2'}(\Lambda_1, \Lambda_2) \\
&\times \bm{\widehat{D}} { \exp  \big[-x_1 u_1 v_1+x_2
u_1+x_3 v_1 -x_4 u_2 v_2 } \\
& +x_5 u_2+x_6 v_2+ x_7  (u_1 v_2+v_1 u_2 ) \big],\\
\end{aligned}
\end{equation}
where the coefficients $x_i$ are as under 
\begin{eqnarray}
&&x_0=(T_C \lambda^2-1)^{-1},\,\quad  x_1=2x_0(1-T_C), \nonumber\\
&&x_2=y_0+x_0 y_2, \quad \quad \quad x_3=y_0+x_0 y_3, \nonumber \\
&&x_4=a_4 \lambda^2, \quad \quad \quad \quad \quad\, x_5=y_1+x_0 y_4, \nonumber \\
&&x_6=y_1+x_0 y_5, \quad \quad \quad x_7=2x_0(\lambda^2-1)\sqrt{T_C},
\end{eqnarray}
with
\begin{equation}\label{eq:y}
\begin{aligned}
y_0=&x_0 d (\lambda +1) \sqrt{1-\lambda ^2} \sqrt{\left(1-T_C\right) T_C},\\
y_1=&-x_0 d (\lambda +1) \sqrt{1-\lambda ^2} \sqrt{1-T_C}.\\
\end{aligned}
\end{equation}
The operator $\bm{\widehat{D}} $ is  given by
\begin{equation}
\begin{aligned}
\bm{\widehat{D}} = \frac{1}{(m!)^2}
\frac{\partial^{m}}{\partial\,u_1^{m}}
\frac{\partial^{m}}{\partial\,v_1^{m}} 
\frac{\partial^{m}}{\partial\,u_2^{m}}
\frac{\partial^{m}}{\partial\,v_2^{m}} \{ \bullet
\}_{\substack{u_1= v_1=0 \\ u_2= v_2=0}}.\\
\end{aligned}
\end{equation}
The success probability of $m$-photon catalysis operation is evaluated  from
Eq.~(\ref{reducedcal}) as
\begin{equation}\label{probps}
\begin{aligned}
P_{PC}^{(m)} &=\widetilde{ \chi}^{(m)}_{A_1 A'_2}(\Lambda_1,
\Lambda_2) \bigg|_{\substack{\tau_1= \sigma_1=0\\ \tau_2=
\sigma_2=0}}\\
& =P_{PC}^{(0)}  \bm{\widehat{D}} { \exp  \big[-x_1 u_1
v_1+y_0 (u_1+ v_1) -x_4 u_2 v_2 } \\
& +y_1 (u_2+ v_2)+ x_7  (u_1 v_2+v_1 u_2 )\big],\\
\end{aligned}
\end{equation}
where the coefficients $y_i$ are given in
Eq.~(\ref{eq:y}). 
Finally, the normalized  Wigner characteristic function   of
the  $m$-PCTMSC state  is given by
\begin{equation}\label{eq:normalized}
\chi^{(m)}_{A_1 A_2'}(\Lambda_1, \Lambda_2) =\left(P_{PS}^{
(m) }\right)^{-1}\widetilde{ \chi}^{(m)}_{A_1
A'_2}(\Lambda_1, \Lambda_2) .
\end{equation}
Wigner characteristic function of PCTMSV states can be
obtained by setting $d=0$ in the above equation. Further, in
the limit $T_C \rightarrow 1$ in Eq.~(\ref{eq:normalized}),
we obtain the Wigner function of the TMSC state.

To evaluate the secret key rate, we need to compute the
covariance matrix of the PCTMSC state~\cite{Cerf-prl-2006}. 
The covariance matrix contains the averages of different
symmetrically ordered operators.
One of the major advantages of working in the Wigner
characteristic formalism is that 
it is possible to determine the average of a symmetrically
ordered operator by differentiating the $m$-PCTMSC state's
Wigner characteristic function with regard to the $\tau$ and
$\sigma$ parameters as
\begin{equation}\label{app:covfinalch}
\begin{aligned}
{}_{\bm{:}}^{\bm{:}}  \hat{q_1}^{r_1} \hat{p_1}^{s_1}
\hat{q_2}^{r_2} \hat{p_2}^{s_2} {}_{\bm{:}}^{\bm{:}}   =
\bm{\widehat{F}} \chi^{(m)}_{A_1 A_2'} (\tau_1,
\sigma_1,\tau_2, \sigma_2) , 
\end{aligned}
\end{equation}
with 
\begin{equation}
\begin{aligned}
\bm{\widehat{F}} =&\left( \frac{1}{i}
\right)^{r_1+r_2}\left( \frac{1}{-i} \right)^{s_1+s_2}
\frac{\partial^{r_1+s_1}}{\partial \sigma_1^{r_1} \partial
\tau_1^{s_1} }\\
&\times\frac{\partial^{r_2+s_2}}{\partial \sigma_2^{r_2}
\partial \tau_2^{s_2} } \{ \bullet  \}_{\substack{\tau_1=
\sigma_1=0\\ \tau_2=
\sigma_2=0}},
\end{aligned}
\end{equation}
and 
the symbol ${}_{\bm{:}}^{\bm{:}}  \bullet
{}_{\bm{:}}^{\bm{:}} $  denotes Weyl ordering.
By selecting appropriate values for $r_1$, $s_1$, $r_2$,
$s_2$ in the   Eq.~(\ref{app:covfinalch}),   all the
elements of the covariance matrix can be obtained. The
covariance matrix takes the following specific form for the
considered PCTMSC state:
\begin{equation}
\Sigma_{A_1 A'_2} = \begin{pmatrix}
V_{A}^{q} & 0 & V_{C}^{q} & 0 \\
0 & V_{A}^{p} & 0 & V_{C}^{p} \\
V_{C}^{q} & 0 & V_{B}^{q} & 0 \\
0 & V_{C}^{p} & 0 & V_{B}^{p}
\end{pmatrix}.
\label{eq:variance_a1a2}
\end{equation}
The covariance matrix for $m$-PCTMSC state  ($m\geq1$)  does
depend on the displacement. 
\section{$m$-PCTMSC state based CV-MDI-QKD protocol}\label{cvmdi}
\begin{figure*} 
\centering
\includegraphics[scale=1]{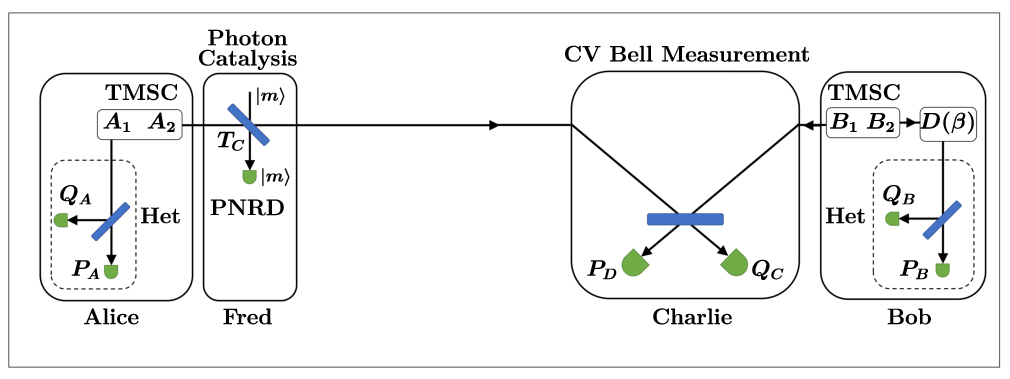}
\caption{    Schematic illustrating the working principle of
CV-MDI-QKD protocol based on the PCTMSC state.
To begin with,  Alice and Bob prepare a TMSC state.   Alice
then transfers one of her modes to an unreliable party,
Fred, to perform PC operation. The photon-catalyzed mode is
then transferred to  another unreliable party, Charlie, who
uses a balanced beam splitter to combine it with the mode
obtained from Bob. Charlie then executes homodyne
measurement on the output modes and publishes the results.
Based on the results, Bob performs a displacement operation,
which entangles the modes   with Alice and Bob. After that,
Alice and Bob measure their modes using heterodyne
measurement.} 
\label{mdi_scheme}
\end{figure*}

We introduce a CV-MDI-QKD protocol that is based on the
$m$-PCTMSC state, with its schematic depicted in
Fig.~\ref{mdi_scheme}. The protocol involves two primary
parties, Alice and Bob, who seek to establish a shared
secure key. To achieve this, they undertake the methodology
described below:

\noindent
{\bf Step 1:} Alice initializes a TMSC state with its two
modes termed as $A_1$ and $A_2$. The variance of the state
$V$, is fixed at $V = \cosh(2r)$.

\noindent
{\bf Step 2:} Alice sends one of the modes (say $A_2$) to
Fred who implements the PC operation. The output
mode($A'_2$) is further transferred to Charlie through a
quantum channel of length $L_{AC}$. This is done to
facilitate entanglement swapping.

\noindent
{\bf Step 3:} Bob generates his own TMSC state with
associated variance $V$. Let the two modes be denoted by
$B_1$ and $B_2$. He then sends one of the modes ($B_1$) to
Charlie via a separate quantum channel with length $L_{BC}$.

\noindent
{\bf Step 4:} Charlie utilizes a beam splitter to combine
the received modes $A'_2$ and $B_1$. Subsequently, he
subjects the resulting output modes C and D, to homodyne
measurements using $\hat{q}$ and $\hat{p}$ operators, and
declares the respective measurement outcomes, $\lbrace Q_C,
P_D\rbrace$.

\noindent
{\bf Step 5:} Based on the results declared by Charlie, Bob
applies a suitable displacement operation to his retained
mode $B_2$, specifically $\hat{D}(g(Q_C+ iP_D))$, with `$g$'
representing the gain factor. This transformation results in
$B_2$ becoming $B'_2$, effectively concluding the
entanglement swapping process. The final state of modes
$A_1$ and $B'_2$ is entangled. Thereafter, Alice and Bob
perform heterodyne measurements on their retained modes to
obtain correlated outcomes, $\lbrace Q_A, P_A\rbrace$ and
$\lbrace Q_B, P_B\rbrace$, respectively.

\noindent
{\bf Step 6:} The final step of CV-MDI-QKD is classical data
post-processing, which involves information reconciliation
(reverse reconciliation) and privacy amplification.


\subsection{Secret key rate of PCTMSC based CV-MDI-QKD}
 We now calculate 
the secret key rate of the PCTMSC-based
CV-MDI-QKD protocol. While the zero PC operation preserves the
Gaussian character of the TMSC state, multi PC operation
transforms the Gaussian state into a non-Gaussian state. We
utilize the covariance matrix of the non-Gaussian state to
calculate the secret key rate. This 
provides a lower bound
  on the secret key rate of the protocol
with non-Gaussian
states as the resource states~\cite{Cerf-prl-2006}.

As the CV-MDI-QKD protocol involves two quantum channels,
Eve has the opportunity to employ two different methods for
eavesdropping. In the first method, Eve carries out a
correlated two-mode coherent Gaussian attack, introducing
quantum correlations into both quantum channels, which is
commonly referred to as the two-mode
attack~\cite{ottaviani-pra-2015,Pirandola-np-2015}.  In the
second method, Eve independently implements entangling
cloner attacks on each quantum channel, called as one-mode
attack~\cite{Pirandola-np-2015}. In our calculation of the
secret key rate, we assume a one-mode collective Gaussian
attack, which is a less effective eavesdropping strategy
employed by Eve.

Since $L_{AC}$ and $L_{BC}$ are the transmission
distances between Alice to Charlie and Bob to Charlie, 
the total
transmission distance in the one-way CV-QKD protocol from
Alice to Bob is $L_{AB}=L_{AC}+L_{BC}$. Here, the distance
shall be expressed in kilometer (km).
The quantum channel typically consists of an optical fiber
network. 
The total loss in the optical fiber  for a length $L$  is
quantified as $\gamma L$, with  $\gamma$ representing the
attenuation factor in decibels per kilometer
(dB/km)~\cite{Grasselli2021}.  In this study, we adopt a
value of $0.2$ dB/km for $\gamma$~\cite{Liao2017}.

Let  the quantum channel between Alice and Charlie be
characterized   via the transmissivity $T_A$ and excess
noise $\varepsilon_A^{\text{th}}$. Similarly, the quantum
channel between Bob and Charlie is characterized   via the
transmissivity $T_B$ and excess noise
$\varepsilon_B^{\text{th}}$. The length and transmissivities
of these quantum channels are related by the relation
$T_A=10^{-\gamma
\frac{L_{AC}}{10}}$ and $T_B=10^{-\gamma
\frac{L_{BC}}{10}}$. 

Under the assumption that all of Bob's operations, except
for the heterodyne detection, are untrustable, the
CV-MDI-QKD protocol under consideration becomes a one-way
QKD protocol that makes use of heterodyne
detection~\cite{oneway,prl2004,Li-pra-2014}.
It is important to highlight that the secret key rate for
this equivalent one-way QKD protocol is either less than or
equal to the original protocol. We employ this one-way QKD
protocol for the sake of simplifying our calculations when
evaluating the secret key rate~\cite{winter}.

The corresponding transmissivity $T$ of the quantum channel
in one-way QKD protocol can be given by $T=\frac{g^2}{2}
T_A$,  where $g$ represents the gain of the displacement
operation implemented by Bob. Similarly, the corresponding
excess noise  $\varepsilon^{\text{th}}$  at optimal gain
turns out to be 
\begin{equation}
\varepsilon^{\text{th}}=\frac{T_B}{T_A}\left(\varepsilon_B^{\text{th}}
-2\right)+
\varepsilon_A^{\text{th}}+\frac{2}{T_A}.
\label{eq:epsilon_thermal}
\end{equation}
Hence, the total channel-added noise turns out to be 
\begin{equation}
\chi_{\text{ch}} = \frac{1-T}{T}+\varepsilon^{\text{th}}.
\end{equation}
Using   $I_{AB}$ to represent  the mutual information
between Alice and
Bob and $\chi_{BE}$ to represent  the Holevo bound between
Bob and Eve,   the secret key rate of the CV-MDI-QKD
protocol  can be expressed as
\begin{equation} 
K=P^{(m)}_{PC}\left(\beta I_{AB}-\chi_{BE}\right).
\label{eq:secure_keyrate}
\end{equation}
Here $\beta$ represents the reconciliation efficiency.  The
calculation details for $I_{AB}$ and $\chi_{BE}$ are
available in Refs.~\cite{Ma-pra-2018,chandan-pra-2019}.
 The central point is that the covariance matrix elements
are used to calculate the key rate  in
Eq.(\ref{eq:secure_keyrate}), which we can compute from
the Wigner characteristic function described in Sec.~\ref{wig:pctmsc}.
 
\section{Numerical    calculation
 and analysis of the secret key
rate}\label{numerical}
Having computed the covariance matrix for the $m$-PCTMSC
state, we now move on to   numerically  calculate
  the secret key rate  for the family of $m$-PCTMSC states.
Our focus will be on the extreme asymmetric
case, when $L_{BC} = 0$, \ie, Charlie and Bob are together
as this scenario yields maximum transmission distance.
Initially, we will discuss the secret key rate dependence on
displacement for specific values of variance and
transmissivity. Later, we will optimize the secret key rate
with respect to PCTMSC state parameters, namely,  variance,
transmissivity, and displacement to obtain a global
picture.
\subsection{The impact of displacement on transmission
distance for PCTMSC states}
\label{sec:contour}
In this section, we first scrutinize the  effect of zero PC
operation, wherein we explain the inherent limitations of
the `0-PCTMSC state' as compared to the `0-PCTMSV state' and
then move on to the non-Gaussian `1-PCTMSC state', where we
show that the right amount of displacement can substantially
augment the performance   for certain parameter
value range.
\subsubsection{Secret key   rate for 0-PCTMSC state}
The covariance matrix of the 0-PCTMSC state is independent
of the displacement, thereby rendering the factor
$\left(\beta I_{AB}-\chi_{BE}\right)$  in the secret key
rate~(\ref{eq:secure_keyrate}) identical for both 0-PCTMSC
and 0-PCTMSV states. However, as shown in
Sec.~\ref{wig:zpc}, the success probability for 0-PCTMSC
state  decreases with increasing displacement. Therefore,
the secret key rate, which is a product of the probability
and  $\left(\beta I_{AB}-\chi_{BE}\right)$,  of the 0-PCTMSC
state is always less than 0-PCTMSV state. This, in turn,
reduces the transmission distance when the displacement of
the TMSC state is increased at a fixed secret key rate.

\begin{figure}[h!] 
\centering
\includegraphics[scale=1]{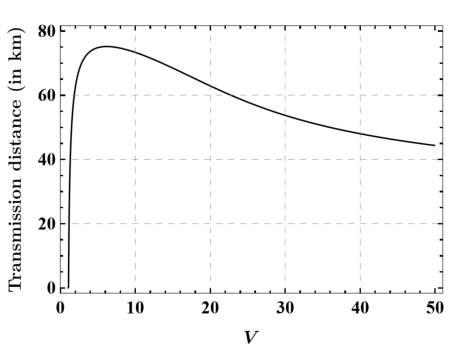}
\caption{ Transmission distance $L_{AB}$ as a function of  
variance $V$  at a fixed secret key rate  $K=10^{-3}$ for
the TMSV state.  
Numerical values of the other parameters are taken as $\beta
= 96\%$ and
$\epsilon_A^{\text{th}}=\epsilon_B^{\text{th}}=0.002$.  }
\label{contourtmsv}
\end{figure}

 To understand how zero photon catalysis may  appear to 
enhance the maximum transmission
distance~\cite{catcvmdiqkd2020}, two facts are important.
Firstly, the fact that the 0-PCTMSV state is essentially
another Gaussian state with a reduced variance as compared
to the input state. Secondly, as presented in
Fig.~\ref{contourtmsv}, if we plot variance as a
function of maximum distance for a given key rate for the
TMSV state, there is an optimal value of variance.  Now, if
the  initial variance of the TMSV state is set higher than
the optimal value, the 0-PC may lead to an 
increased value of the
maximum transmission distance as it reduces the variance.
This, however, is an artifact of working with a non-optimal
value of variance and not a real advantage of zero photon
catalysis.  This is precisely the reason for the results
shown in Ref.~\cite{catcvmdiqkd2020}, a TMSV state with a
variance of $V=15$ was considered, which revealed that zero
photon catalysis could indeed enhance the maximum
transmission distance. However, our findings, as illustrated
in Fig.~\ref{partial}(c), indicate that when dealing with a
TMSV state with smaller variance ($V=5$), zero photon
catalysis does not lead to an increase in the transmission
distance.
The conclusion is that 0-PC is of no real
use. In fact once it becomes clear that after 0-PC, the state is still a Gaussian state with
deteriorated parameters, there is no question of this state
giving any advantage over the initial state in any quantum
protocol. 

There are several problems with the results presented  in
Ref.~\cite{catalysis2021} in the context of zero photon
catalysis; firstly, they have made a wrong conclusion that
after zero photon catalysis the state becomes non-Gaussian;
secondly, their graphs and text contain discrepancies
on the role of displacement. 
\subsubsection{Secret key rate for 1-PCTMSC state}
\begin{figure}[h!] 
\centering
\includegraphics[scale=1]{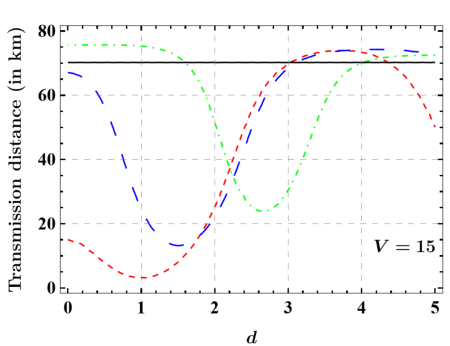}
\caption{Transmission distance $L_{AB}$ as a function of
displacement $d$ for different beam splitter
transmissivities  for the 1-PCTMSC state.  The secret key
rate is fixed at $K=10^{-5}$. Different curves for  1-PCTMSC
state correspond to $T_C = 0.90$ (dotted), $T_C = 0.94$ (
dashed), and $T_C = 0.98$ (dotdashed). The solid straight
line represents the TMSV state.  Numerical values of the
other parameters are taken as $\beta = 96\%$ and
$\epsilon_A^{\text{th}}=\epsilon_B^{\text{th}}=0.002$.
}
\label{contour}
\end{figure}

We now turn to the analysis of 1-PCTMSC state and show
that 1-PC operation on TMSV and TMSC states can have
positive effects on the QKD performance.  We first note that
the performance of $1$-PCTMSC state  is contingent on
displacement since the covariance matrix depends on
displacement.  We analyze the variation in the maximum
transmission distance as a function of displacement  for
1-PCTMSC state at fixed secret key rate.  The results are
shown in Fig.~\ref{contour}, where we have considered a
constant variance $V=15$ and three different values of
transmissivities  $T_C=0.90$, $0.94$, and $0.98$ at a fixed
key rate $K=10^{-5}$. 

The results show that 1-PC operation can have advantageous
effects on TMSV state. Further, introducing displacement can
lead to either  beneficial or harmful outcomes depending on
the   transmissivity value. For instance, in the case of
$T_C=0.90$,  introducing displacement in the range of $d \in
(3,4)$ yields optimal performance, outperforming   the TMSV
state. In Ref.~\cite{catalysis2021} (Fig.~9), 1-PCTMSC state
is shown to yield lower performance than the TMSV state at
$d=2$, $V=50$, and $T_C=0.90$ .  This is even evident in
our result at $V=15$ for the curve $T_C=0.90$, where the
performance is quite low at $d=2$. Therefore, introducing
the right amount of displacement is crucial to achieving
optimal performance.
Furthermore, introducing displacement in the range $d \in
(3.5,4.5)$ is beneficial while working at $T_C=0.94$. 
Finally, at $T_C=0.98$,   displacement does not  help in
improving the transmission distance.
\begin{figure}[h!] 
\centering
\includegraphics[scale=1]{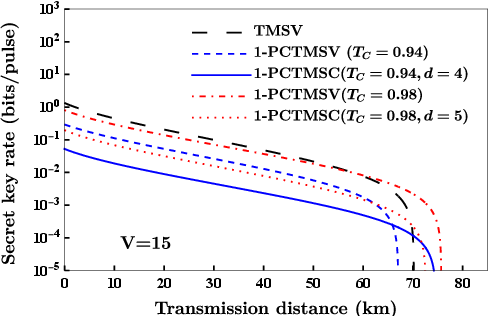}
\caption{ Secret key rate  as a function of transmission
distance. Numerical values of the other parameters are taken
as $\beta = 96\%$ and
$\epsilon_A^{\text{th}}=\epsilon_B^{\text{th}}=0.002$. }
\label{onedfixed}
\end{figure}

These results can be explicitly seen in
Fig.~\ref{onedfixed}, where we plot the secret key rate as a
function of transmission distance at fixed transmissivities
and variance. At $T_C=0.94$, the maximum transmission
distance of the 1-PCTMSV state is less than the TMSV state,
while the maximum transmission distance of the 1-PCTMSC
state is more than the TMSV state. On the other hand, at
$T_C=0.98$, the maximum transmission distance for both
1-PCTMSV and 1-PCTMSC states is more than than the TMSV
state; however, the maximum transmission distance of the
1-PCTMSV state is more than the 1-PCTMSC state.

We would like to point out that our results for 1-PC
states like those for 0-PC state are at variance with
the conclusions drawn in Ref.~\cite{catalysis2021}.
Secondly, as will become clear from the results in the next section
and as was seen in the previous section, the advantage of
displacement and PC are primarily in the high variance
regime which is not the optimal value of variance.
\subsection{Optimization of the secret key rate with state
parameters }\label{sec:opt}
 
The state parameters for PCTMSC states consist of variance
which represents the amount of squeezing, displacement which
represents the amplification factor, and transmissivity which
represents the strength of the PC process.  For the PCTMSV
state, where displacement is set to zero, the state
parameters are variance and transmissivity only. 
In an actual practical experimental situation where the 
experimentalist can generate a state with varying
variance,  displacement, and  transmissivity,  it is very
important to find out what is the best of choice of these
parameters for carrying out the QKD protocol.
In such a scenario, we are interested in
finding out the optimal value of parameters that maximizes
the secret key rate of the QKD protocol.
Two different situations are
considered for the optimization of the secret key rate:
\begin{itemize}
\item  Variance is kept fixed and the key rate is optimized
with respect to displacement and transmissivity and cases of
no PC, 0 PC and 1 PC. 
 
\item  The key rate is optimized with respect to all the three
parameters namely variance, displacement and transmissivity
for the no PC 0 PC  and 1 PC cases. 
\end{itemize}

\subsubsection{Optimal displacement and transmissivity for
fixed variance}
  We maximize the secret key rate by
 varying the displacement and transmissivity at a constant variance.
We chose three difference values of the variance
parameter namely,
$V = 5$, $V=10$, and $V=15$. The displacement parameter is
restricted to the interval of $d \in (0,5)$.  The
optimization problem can be succinctly expressed as
\begin{equation}
\begin{aligned}
\max_{d, T_C} \quad & K(V,d,T_C)\\
\textrm{subject to} \quad & 0\leq d \leq 5,\\
&0\leq T_C \leq 1.  \\
\end{aligned}
\end{equation}

\begin{figure}[h!] 
\centering
\includegraphics[scale=1]{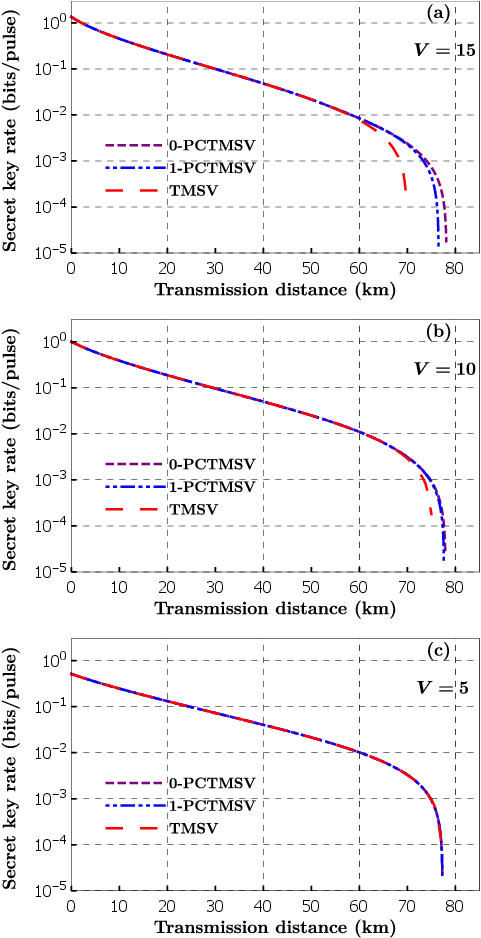}
\caption{Secret key rate  (optimized with respect to
displacement  and transmissivity) as a function of
transmission distance. Three different variances have been
considered: (a) $V = 5$, (b) $V = 10$, (c) $V = 15$.   
Numerical values of the other parameters are taken as $\beta
= 96\%$ and
$\epsilon_A^{\text{th}}=\epsilon_B^{\text{th}}=0.002$.  }
\label{partial}
\end{figure}
The optimization of state parameters is carried out
for each transmission distance value. We only need to
optimize the transmissivity parameter  to achieve maximum
secret key rate for the PCTMSV state.  As expected, the
optimal displacement for the 
0-PCTMSC state turns out to be zero. Surprisingly,  
the optimal displacement for the 1-PCTMSC state is also
zero. Since the optimal performance of the  0-PCTMSC and
1-PCTMSC states  is the same as that of 0-PCTMSV and
1-PCTMSV states, respectively, we present the results only
for   0-PCTMSV  and 1-PCTMSV states along with the TMSV
state in Fig.~\ref{partial} for three fixed values of
variance   $(V = 5$, $10$, and $15)$.

At high variance ($V=15$), both 0-PCTMSV and 1-PCTMSV states
outperform the TMSV state.  The maximum transmission
distance is obtained for the 0-PCTMSV state followed by the
1-PCTMSV state. We note that this result for the 0-PCTMSV
state has already been obtained in a previous
study~\cite{catcvmdiqkd2020}.  As   the variance is
decreased, all three states tend to yield the same
performance.  In particular, at $V=5$, we observe that the
performance of the three states is similar. However, we also
notice a decrease in the secret key rate as the variance
decreases.  It should be noted that the optimal
transmissivity approaches unity wherever the PCTMSV states
performance is equal to the TMSV state.  This stems from the
fact that the PCTMSV states approach the TMSV state in the
unit transmissivity limit.

This leaves us with the task of finding the optimal variance
of the employed resource states. The next section attempts
to determine the optimal variance maximizing the secret key
rate for different values of transmission distance. 
\subsubsection{Optimal variance, displacement and transmissivity}
We are now ready to we carry out the optimization
of all state parameters, namely, variance, displacement, and
transmissivity,  with the aim of maximizing the secret key
rate. The range of these parameters considered in the
optimization is following:
\begin{equation}
\begin{aligned}
\max_{V,d, T_C} \quad & K(V,d,T_C)\\
\textrm{subject to} \quad & 1\leq V \leq 15,\\
& 0\leq d \leq 5,\\
&0\leq T_C \leq 1.  \\
\end{aligned}
\end{equation}
\begin{figure}[h!] 
\centering
\includegraphics[scale=1]{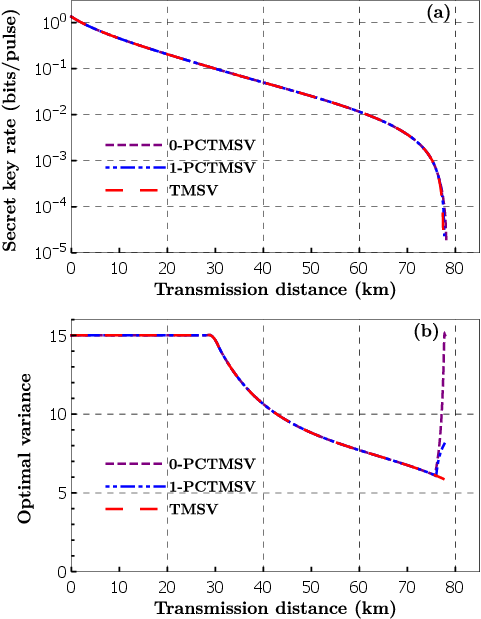}
\caption{(a) Secret key rate  (optimized with respect to
variance, displacement,  and transmissivity) as a function
of the transmission distance.  (b) 
Optimal variance as a function of the transmission distance.
Numerical values of the other parameters are taken as $\beta
= 96\%$ and
$\epsilon_A^{\text{th}}=\epsilon_B^{\text{th}}=0.002$.}
\label{all}
\end{figure}
The result of the optimization is shown in Fig.~\ref{all}.
We see that the PC operation can enhance the maximum
transmission distance by less than one km. Further, the
optimal variance is $V\approx 6$ at the maximum transmission
distance; however, the optimal variance rises sharply for
the 0-PCTMSV state.  As states with low variances are easy
to prepare, it raises a big question on the utility of
photon catalysis by Alice in CV-MDI-QKD. Our analysis
demonstrates that there is very little advantage in
performance if Alice  performs PC operations.
\section{Conclusion}
\label{sec:conc}
In this article, we considered multiphoton catalysis on the
TMSC state (the TMSV state is a special case). The generated
state is used as a resource for CV-MDI-QKD.
We showed that Alice's implementation of zero and single
photon catalysis has a negligible impact on enhancing the
maximum transmission distance. Additionally, the results
demonstrate that displacement does not yield any advantage
in photon catalysis.
The results provide a global perspective as we optimize
all the state parameters and thereby show that many
earlier results were artifacts of working in the wrong
parameter regime. We presented our results stepwise so that
we can make comparisons with earlier work.

A key ingredient of our work is that we computed the Wigner
characteristic function of the PCTMSC states, which can be
utilized to examine non-Gaussianity~\cite{non-G},
nonlocality~\cite{am2002,nonlocality}, and
nonclassicality~\cite{Hillery,Mandel:79,antibunching}. In an
earlier work~\cite{long-qkd}, we demonstrated that applying
photon subtraction on Alice's end does not yield any
benefits in CV-MDI-QKD. In future research, we aim to
investigate whether noiseless linear
amplifiers~\cite{nlamdi} and quantum scissors~\cite{s-mdi}
can enhance the performance of CV-MDI-QKD.

Reference~\cite{catalysis2021} claims that
employing non-Gaussian operations on Bob's side results in a
final state with reduced entanglement content, potentially
leading to a ``lesser key rate or transmission distance".
However, Ref.~\cite{betterresource} has   demonstrated that
entanglement content may be enhanced by performing
non-Gaussian operation by Bob (after evolution through the
noisy channel) as compared to non-Gaussian operation
performed by Alice. Therefore, it is worthwhile to explore
whether performing non-Gaussian operations on Bob's side
offers more advantages than doing so on Alice's side.
\section*{Acknowledgement}
A and C.K. acknowledge  the financial
support from {\bf DST/ICPS/QuST/Theme-1/2019/General} Project
number {\sf Q-68}. 

\begin{thebibliography}{55}%
	\makeatletter
	\providecommand \@ifxundefined [1]{%
		\@ifx{#1\undefined}
	}%
	\providecommand \@ifnum [1]{%
		\ifnum #1\expandafter \@firstoftwo
		\else \expandafter \@secondoftwo
		\fi
	}%
	\providecommand \@ifx [1]{%
		\ifx #1\expandafter \@firstoftwo
		\else \expandafter \@secondoftwo
		\fi
	}%
	\providecommand \natexlab [1]{#1}%
	\providecommand \enquote  [1]{``#1''}%
	\providecommand \bibnamefont  [1]{#1}%
	\providecommand \bibfnamefont [1]{#1}%
	\providecommand \citenamefont [1]{#1}%
	\providecommand \href@noop [0]{\@secondoftwo}%
	\providecommand \href [0]{\begingroup \@sanitize@url \@href}%
	\providecommand \@href[1]{\@@startlink{#1}\@@href}%
	\providecommand \@@href[1]{\endgroup#1\@@endlink}%
	\providecommand \@sanitize@url [0]{\catcode `\\12\catcode `\$12\catcode
		`\&12\catcode `\#12\catcode `\^12\catcode `\_12\catcode `\%12\relax}%
	\providecommand \@@startlink[1]{}%
	\providecommand \@@endlink[0]{}%
	\providecommand \url  [0]{\begingroup\@sanitize@url \@url }%
	\providecommand \@url [1]{\endgroup\@href {#1}{\urlprefix }}%
	\providecommand \urlprefix  [0]{URL }%
	\providecommand \Eprint [0]{\href }%
	\providecommand \doibase [0]{https://doi.org/}%
	\providecommand \selectlanguage [0]{\@gobble}%
	\providecommand \bibinfo  [0]{\@secondoftwo}%
	\providecommand \bibfield  [0]{\@secondoftwo}%
	\providecommand \translation [1]{[#1]}%
	\providecommand \BibitemOpen [0]{}%
	\providecommand \bibitemStop [0]{}%
	\providecommand \bibitemNoStop [0]{.\EOS\space}%
	\providecommand \EOS [0]{\spacefactor3000\relax}%
	\providecommand \BibitemShut  [1]{\csname bibitem#1\endcsname}%
	\let\auto@bib@innerbib\@empty
	\bibitem [{\citenamefont {Gisin}\ \emph {et~al.}(2002)\citenamefont {Gisin},
		\citenamefont {Ribordy}, \citenamefont {Tittel},\ and\ \citenamefont
		{Zbinden}}]{gisin-rep-2002}%
	\BibitemOpen
	\bibfield  {author} {\bibinfo {author} {\bibfnamefont {N.}~\bibnamefont
			{Gisin}}, \bibinfo {author} {\bibfnamefont {G.}~\bibnamefont {Ribordy}},
		\bibinfo {author} {\bibfnamefont {W.}~\bibnamefont {Tittel}},\ and\ \bibinfo
		{author} {\bibfnamefont {H.}~\bibnamefont {Zbinden}},\ }\bibfield  {title}
	{\bibinfo {title} {Quantum cryptography},\ }\href
	{https://doi.org/10.1103/RevModPhys.74.145} {\bibfield  {journal} {\bibinfo
			{journal} {Rev. Mod. Phys.}\ }\textbf {\bibinfo {volume} {74}},\ \bibinfo
		{pages} {145} (\bibinfo {year} {2002})}\BibitemShut {NoStop}%
	\bibitem [{\citenamefont {Scarani}\ \emph {et~al.}(2009)\citenamefont
		{Scarani}, \citenamefont {Bechmann-Pasquinucci}, \citenamefont {Cerf},
		\citenamefont {Du\ifmmode~\check{s}\else \v{s}\fi{}ek}, \citenamefont
		{L\"utkenhaus},\ and\ \citenamefont {Peev}}]{scarani-rmp-2009}%
	\BibitemOpen
	\bibfield  {author} {\bibinfo {author} {\bibfnamefont {V.}~\bibnamefont
			{Scarani}}, \bibinfo {author} {\bibfnamefont {H.}~\bibnamefont
			{Bechmann-Pasquinucci}}, \bibinfo {author} {\bibfnamefont {N.~J.}\
			\bibnamefont {Cerf}}, \bibinfo {author} {\bibfnamefont {M.}~\bibnamefont
			{Du\ifmmode~\check{s}\else \v{s}\fi{}ek}}, \bibinfo {author} {\bibfnamefont
			{N.}~\bibnamefont {L\"utkenhaus}},\ and\ \bibinfo {author} {\bibfnamefont
			{M.}~\bibnamefont {Peev}},\ }\bibfield  {title} {\bibinfo {title} {The
			security of practical quantum key distribution},\ }\href
	{https://doi.org/10.1103/RevModPhys.81.1301} {\bibfield  {journal} {\bibinfo
			{journal} {Rev. Mod. Phys.}\ }\textbf {\bibinfo {volume} {81}},\ \bibinfo
		{pages} {1301} (\bibinfo {year} {2009})}\BibitemShut {NoStop}%
	\bibitem [{\citenamefont {Pirandola}\ \emph {et~al.}(2020)\citenamefont
		{Pirandola}, \citenamefont {Andersen}, \citenamefont {Banchi}, \citenamefont
		{Berta}, \citenamefont {Bunandar}, \citenamefont {Colbeck}, \citenamefont
		{Englund}, \citenamefont {Gehring}, \citenamefont {Lupo}, \citenamefont
		{Ottaviani}, \citenamefont {Pereira}, \citenamefont {Razavi}, \citenamefont
		{Shaari}, \citenamefont {Tomamichel}, \citenamefont {Usenko}, \citenamefont
		{Vallone}, \citenamefont {Villoresi},\ and\ \citenamefont
		{Wallden}}]{Pirandola-2019}%
	\BibitemOpen
	\bibfield  {author} {\bibinfo {author} {\bibfnamefont {S.}~\bibnamefont
			{Pirandola}}, \bibinfo {author} {\bibfnamefont {U.~L.}\ \bibnamefont
			{Andersen}}, \bibinfo {author} {\bibfnamefont {L.}~\bibnamefont {Banchi}},
		\bibinfo {author} {\bibfnamefont {M.}~\bibnamefont {Berta}}, \bibinfo
		{author} {\bibfnamefont {D.}~\bibnamefont {Bunandar}}, \bibinfo {author}
		{\bibfnamefont {R.}~\bibnamefont {Colbeck}}, \bibinfo {author} {\bibfnamefont
			{D.}~\bibnamefont {Englund}}, \bibinfo {author} {\bibfnamefont
			{T.}~\bibnamefont {Gehring}}, \bibinfo {author} {\bibfnamefont
			{C.}~\bibnamefont {Lupo}}, \bibinfo {author} {\bibfnamefont {C.}~\bibnamefont
			{Ottaviani}}, \bibinfo {author} {\bibfnamefont {J.~L.}\ \bibnamefont
			{Pereira}}, \bibinfo {author} {\bibfnamefont {M.}~\bibnamefont {Razavi}},
		\bibinfo {author} {\bibfnamefont {J.~S.}\ \bibnamefont {Shaari}}, \bibinfo
		{author} {\bibfnamefont {M.}~\bibnamefont {Tomamichel}}, \bibinfo {author}
		{\bibfnamefont {V.~C.}\ \bibnamefont {Usenko}}, \bibinfo {author}
		{\bibfnamefont {G.}~\bibnamefont {Vallone}}, \bibinfo {author} {\bibfnamefont
			{P.}~\bibnamefont {Villoresi}},\ and\ \bibinfo {author} {\bibfnamefont
			{P.}~\bibnamefont {Wallden}},\ }\bibfield  {title} {\bibinfo {title}
		{Advances in quantum cryptography},\ }\href
	{https://doi.org/10.1364/AOP.361502} {\bibfield  {journal} {\bibinfo
			{journal} {Adv. Opt. Photon.}\ }\textbf {\bibinfo {volume} {12}},\ \bibinfo
		{pages} {1012} (\bibinfo {year} {2020})}\BibitemShut {NoStop}%
	\bibitem [{\citenamefont {Braunstein}\ and\ \citenamefont
		{Pirandola}(2012)}]{Braunstein-prl-2012}%
	\BibitemOpen
	\bibfield  {author} {\bibinfo {author} {\bibfnamefont {S.~L.}\ \bibnamefont
			{Braunstein}}\ and\ \bibinfo {author} {\bibfnamefont {S.}~\bibnamefont
			{Pirandola}},\ }\bibfield  {title} {\bibinfo {title} {Side-channel-free
			quantum key distribution},\ }\href
	{https://doi.org/10.1103/PhysRevLett.108.130502} {\bibfield  {journal}
		{\bibinfo  {journal} {Phys. Rev. Lett.}\ }\textbf {\bibinfo {volume} {108}},\
		\bibinfo {pages} {130502} (\bibinfo {year} {2012})}\BibitemShut {NoStop}%
	\bibitem [{\citenamefont {Lo}\ \emph {et~al.}(2012)\citenamefont {Lo},
		\citenamefont {Curty},\ and\ \citenamefont {Qi}}]{Lo-prl-2012}%
	\BibitemOpen
	\bibfield  {author} {\bibinfo {author} {\bibfnamefont {H.-K.}\ \bibnamefont
			{Lo}}, \bibinfo {author} {\bibfnamefont {M.}~\bibnamefont {Curty}},\ and\
		\bibinfo {author} {\bibfnamefont {B.}~\bibnamefont {Qi}},\ }\bibfield
	{title} {\bibinfo {title} {Measurement-device-independent quantum key
			distribution},\ }\href {https://doi.org/10.1103/PhysRevLett.108.130503}
	{\bibfield  {journal} {\bibinfo  {journal} {Phys. Rev. Lett.}\ }\textbf
		{\bibinfo {volume} {108}},\ \bibinfo {pages} {130503} (\bibinfo {year}
		{2012})}\BibitemShut {NoStop}%
	\bibitem [{\citenamefont {Pirandola}\ \emph {et~al.}(2015)\citenamefont
		{Pirandola}, \citenamefont {Ottaviani}, \citenamefont {Spedalieri},
		\citenamefont {Weedbrook}, \citenamefont {Braunstein}, \citenamefont {Lloyd},
		\citenamefont {Gehring}, \citenamefont {Jacobsen},\ and\ \citenamefont
		{Andersen}}]{Pirandola-np-2015}%
	\BibitemOpen
	\bibfield  {author} {\bibinfo {author} {\bibfnamefont {S.}~\bibnamefont
			{Pirandola}}, \bibinfo {author} {\bibfnamefont {C.}~\bibnamefont
			{Ottaviani}}, \bibinfo {author} {\bibfnamefont {G.}~\bibnamefont
			{Spedalieri}}, \bibinfo {author} {\bibfnamefont {C.}~\bibnamefont
			{Weedbrook}}, \bibinfo {author} {\bibfnamefont {S.~L.}\ \bibnamefont
			{Braunstein}}, \bibinfo {author} {\bibfnamefont {S.}~\bibnamefont {Lloyd}},
		\bibinfo {author} {\bibfnamefont {T.}~\bibnamefont {Gehring}}, \bibinfo
		{author} {\bibfnamefont {C.~S.}\ \bibnamefont {Jacobsen}},\ and\ \bibinfo
		{author} {\bibfnamefont {U.~L.}\ \bibnamefont {Andersen}},\ }\bibfield
	{title} {\bibinfo {title} {High-rate measurement-device-independent quantum
			cryptography},\ }\href {https://doi.org/10.1038/nphoton.2015.83} {\bibfield
		{journal} {\bibinfo  {journal} {Nature Photonics}\ }\textbf {\bibinfo
			{volume} {9}},\ \bibinfo {pages} {397} (\bibinfo {year} {2015})}\BibitemShut
	{NoStop}%
	\bibitem [{\citenamefont {Li}\ \emph {et~al.}(2014)\citenamefont {Li},
		\citenamefont {Zhang}, \citenamefont {Xu}, \citenamefont {Peng},\ and\
		\citenamefont {Guo}}]{Li-pra-2014}%
	\BibitemOpen
	\bibfield  {author} {\bibinfo {author} {\bibfnamefont {Z.}~\bibnamefont
			{Li}}, \bibinfo {author} {\bibfnamefont {Y.-C.}\ \bibnamefont {Zhang}},
		\bibinfo {author} {\bibfnamefont {F.}~\bibnamefont {Xu}}, \bibinfo {author}
		{\bibfnamefont {X.}~\bibnamefont {Peng}},\ and\ \bibinfo {author}
		{\bibfnamefont {H.}~\bibnamefont {Guo}},\ }\bibfield  {title} {\bibinfo
		{title} {Continuous-variable measurement-device-independent quantum key
			distribution},\ }\href {https://doi.org/10.1103/PhysRevA.89.052301}
	{\bibfield  {journal} {\bibinfo  {journal} {Phys. Rev. A}\ }\textbf {\bibinfo
			{volume} {89}},\ \bibinfo {pages} {052301} (\bibinfo {year}
		{2014})}\BibitemShut {NoStop}%
	\bibitem [{\citenamefont {Ma}\ \emph {et~al.}(2014)\citenamefont {Ma},
		\citenamefont {Sun}, \citenamefont {Jiang}, \citenamefont {Gui},\ and\
		\citenamefont {Liang}}]{Xiang-pra-2014}%
	\BibitemOpen
	\bibfield  {author} {\bibinfo {author} {\bibfnamefont {X.-C.}\ \bibnamefont
			{Ma}}, \bibinfo {author} {\bibfnamefont {S.-H.}\ \bibnamefont {Sun}},
		\bibinfo {author} {\bibfnamefont {M.-S.}\ \bibnamefont {Jiang}}, \bibinfo
		{author} {\bibfnamefont {M.}~\bibnamefont {Gui}},\ and\ \bibinfo {author}
		{\bibfnamefont {L.-M.}\ \bibnamefont {Liang}},\ }\bibfield  {title} {\bibinfo
		{title} {Gaussian-modulated coherent-state measurement-device-independent
			quantum key distribution},\ }\href
	{https://doi.org/10.1103/PhysRevA.89.042335} {\bibfield  {journal} {\bibinfo
			{journal} {Phys. Rev. A}\ }\textbf {\bibinfo {volume} {89}},\ \bibinfo
		{pages} {042335} (\bibinfo {year} {2014})}\BibitemShut {NoStop}%
	\bibitem [{\citenamefont {Opatrn\'y}\ \emph {et~al.}(2000)\citenamefont
		{Opatrn\'y}, \citenamefont {Kurizki},\ and\ \citenamefont
		{Welsch}}]{tel2000}%
	\BibitemOpen
	\bibfield  {author} {\bibinfo {author} {\bibfnamefont {T.}~\bibnamefont
			{Opatrn\'y}}, \bibinfo {author} {\bibfnamefont {G.}~\bibnamefont {Kurizki}},\
		and\ \bibinfo {author} {\bibfnamefont {D.-G.}\ \bibnamefont {Welsch}},\
	}\bibfield  {title} {\bibinfo {title} {Improvement on teleportation of
			continuous variables by photon subtraction via conditional measurement},\
	}\href {https://doi.org/10.1103/PhysRevA.61.032302} {\bibfield  {journal}
		{\bibinfo  {journal} {Phys. Rev. A}\ }\textbf {\bibinfo {volume} {61}},\
		\bibinfo {pages} {032302} (\bibinfo {year} {2000})}\BibitemShut {NoStop}%
	\bibitem [{\citenamefont {Dell'Anno}\ \emph {et~al.}(2007)\citenamefont
		{Dell'Anno}, \citenamefont {De~Siena}, \citenamefont {Albano},\ and\
		\citenamefont {Illuminati}}]{dellanno-2007}%
	\BibitemOpen
	\bibfield  {author} {\bibinfo {author} {\bibfnamefont {F.}~\bibnamefont
			{Dell'Anno}}, \bibinfo {author} {\bibfnamefont {S.}~\bibnamefont {De~Siena}},
		\bibinfo {author} {\bibfnamefont {L.}~\bibnamefont {Albano}},\ and\ \bibinfo
		{author} {\bibfnamefont {F.}~\bibnamefont {Illuminati}},\ }\bibfield  {title}
	{\bibinfo {title} {Continuous-variable quantum teleportation with
			non-gaussian resources},\ }\href {https://doi.org/10.1103/PhysRevA.76.022301}
	{\bibfield  {journal} {\bibinfo  {journal} {Phys. Rev. A}\ }\textbf {\bibinfo
			{volume} {76}},\ \bibinfo {pages} {022301} (\bibinfo {year}
		{2007})}\BibitemShut {NoStop}%
	\bibitem [{\citenamefont {Yang}\ and\ \citenamefont {Li}(2009)}]{tel2009}%
	\BibitemOpen
	\bibfield  {author} {\bibinfo {author} {\bibfnamefont {Y.}~\bibnamefont
			{Yang}}\ and\ \bibinfo {author} {\bibfnamefont {F.-L.}\ \bibnamefont {Li}},\
	}\bibfield  {title} {\bibinfo {title} {Entanglement properties of
			non-gaussian resources generated via photon subtraction and addition and
			continuous-variable quantum-teleportation improvement},\ }\href
	{https://doi.org/10.1103/PhysRevA.80.022315} {\bibfield  {journal} {\bibinfo
			{journal} {Phys. Rev. A}\ }\textbf {\bibinfo {volume} {80}},\ \bibinfo
		{pages} {022315} (\bibinfo {year} {2009})}\BibitemShut {NoStop}%
	\bibitem [{\citenamefont {Xu}(2015)}]{catalysis15}%
	\BibitemOpen
	\bibfield  {author} {\bibinfo {author} {\bibfnamefont {X.-x.}\ \bibnamefont
			{Xu}},\ }\bibfield  {title} {\bibinfo {title} {Enhancing quantum entanglement
			and quantum teleportation for two-mode squeezed vacuum state by local
			quantum-optical catalysis},\ }\href
	{https://doi.org/10.1103/PhysRevA.92.012318} {\bibfield  {journal} {\bibinfo
			{journal} {Phys. Rev. A}\ }\textbf {\bibinfo {volume} {92}},\ \bibinfo
		{pages} {012318} (\bibinfo {year} {2015})}\BibitemShut {NoStop}%
	\bibitem [{\citenamefont {Hu}\ \emph {et~al.}(2017)\citenamefont {Hu},
		\citenamefont {Liao},\ and\ \citenamefont {Zubairy}}]{catalysis17}%
	\BibitemOpen
	\bibfield  {author} {\bibinfo {author} {\bibfnamefont {L.}~\bibnamefont
			{Hu}}, \bibinfo {author} {\bibfnamefont {Z.}~\bibnamefont {Liao}},\ and\
		\bibinfo {author} {\bibfnamefont {M.~S.}\ \bibnamefont {Zubairy}},\
	}\bibfield  {title} {\bibinfo {title} {Continuous-variable entanglement via
			multiphoton catalysis},\ }\href {https://doi.org/10.1103/PhysRevA.95.012310}
	{\bibfield  {journal} {\bibinfo  {journal} {Phys. Rev. A}\ }\textbf {\bibinfo
			{volume} {95}},\ \bibinfo {pages} {012310} (\bibinfo {year}
		{2017})}\BibitemShut {NoStop}%
	\bibitem [{\citenamefont {Wang}\ \emph {et~al.}(2015)\citenamefont {Wang},
		\citenamefont {Hou}, \citenamefont {Chen},\ and\ \citenamefont
		{Xu}}]{wang2015}%
	\BibitemOpen
	\bibfield  {author} {\bibinfo {author} {\bibfnamefont {S.}~\bibnamefont
			{Wang}}, \bibinfo {author} {\bibfnamefont {L.-L.}\ \bibnamefont {Hou}},
		\bibinfo {author} {\bibfnamefont {X.-F.}\ \bibnamefont {Chen}},\ and\
		\bibinfo {author} {\bibfnamefont {X.-F.}\ \bibnamefont {Xu}},\ }\bibfield
	{title} {\bibinfo {title} {Continuous-variable quantum teleportation with
			non-gaussian entangled states generated via multiple-photon subtraction and
			addition},\ }\href {https://doi.org/10.1103/PhysRevA.91.063832} {\bibfield
		{journal} {\bibinfo  {journal} {Phys. Rev. A}\ }\textbf {\bibinfo {volume}
			{91}},\ \bibinfo {pages} {063832} (\bibinfo {year} {2015})}\BibitemShut
	{NoStop}%
	\bibitem [{\citenamefont {Kumar}\ and\ \citenamefont
		{Arora}(2023)}]{tele-2023}%
	\BibitemOpen
	\bibfield  {author} {\bibinfo {author} {\bibfnamefont {C.}~\bibnamefont
			{Kumar}}\ and\ \bibinfo {author} {\bibfnamefont {S.}~\bibnamefont {Arora}},\
	}\bibfield  {title} {\bibinfo {title} {Success probability and performance
			optimization in non-gaussian continuous-variable quantum teleportation},\
	}\href {https://doi.org/10.1103/PhysRevA.107.012418} {\bibfield  {journal}
		{\bibinfo  {journal} {Phys. Rev. A}\ }\textbf {\bibinfo {volume} {107}},\
		\bibinfo {pages} {012418} (\bibinfo {year} {2023})}\BibitemShut {NoStop}%
	\bibitem [{\citenamefont {Kumar}\ \emph
		{et~al.}(2024{\natexlab{a}})\citenamefont {Kumar}, \citenamefont {Sharma},\
		and\ \citenamefont {Arora}}]{noisytele}%
	\BibitemOpen
	\bibfield  {author} {\bibinfo {author} {\bibfnamefont {C.}~\bibnamefont
			{Kumar}}, \bibinfo {author} {\bibfnamefont {M.}~\bibnamefont {Sharma}},\ and\
		\bibinfo {author} {\bibfnamefont {S.}~\bibnamefont {Arora}},\ }\bibfield
	{title} {\bibinfo {title} {Continuous variable quantum teleportation in a
			dissipative environment: Comparison of non-gaussian operations before and
			after noisy channel},\ }\href
	{https://doi.org/https://doi.org/10.1002/qute.202300344} {\bibfield
		{journal} {\bibinfo  {journal} {Advanced Quantum Technologies}\ }\textbf
		{\bibinfo {volume} {7}},\ \bibinfo {pages} {2300344} (\bibinfo {year}
		{2024}{\natexlab{a}})}\BibitemShut {NoStop}%
	\bibitem [{\citenamefont {Birrittella}\ \emph {et~al.}(2012)\citenamefont
		{Birrittella}, \citenamefont {Mimih},\ and\ \citenamefont
		{Gerry}}]{gerryc-pra-2012}%
	\BibitemOpen
	\bibfield  {author} {\bibinfo {author} {\bibfnamefont {R.}~\bibnamefont
			{Birrittella}}, \bibinfo {author} {\bibfnamefont {J.}~\bibnamefont {Mimih}},\
		and\ \bibinfo {author} {\bibfnamefont {C.~C.}\ \bibnamefont {Gerry}},\
	}\bibfield  {title} {\bibinfo {title} {Multiphoton quantum interference at a
			beam splitter and the approach to heisenberg-limited interferometry},\ }\href
	{https://doi.org/10.1103/PhysRevA.86.063828} {\bibfield  {journal} {\bibinfo
			{journal} {Phys. Rev. A}\ }\textbf {\bibinfo {volume} {86}},\ \bibinfo
		{pages} {063828} (\bibinfo {year} {2012})}\BibitemShut {NoStop}%
	\bibitem [{\citenamefont {Carranza}\ and\ \citenamefont
		{Gerry}(2012)}]{josab-2012}%
	\BibitemOpen
	\bibfield  {author} {\bibinfo {author} {\bibfnamefont {R.}~\bibnamefont
			{Carranza}}\ and\ \bibinfo {author} {\bibfnamefont {C.~C.}\ \bibnamefont
			{Gerry}},\ }\bibfield  {title} {\bibinfo {title} {Photon-subtracted two-mode
			squeezed vacuum states and applications to quantum optical interferometry},\
	}\href {https://doi.org/10.1364/JOSAB.29.002581} {\bibfield  {journal}
		{\bibinfo  {journal} {J. Opt. Soc. Am. B}\ }\textbf {\bibinfo {volume}
			{29}},\ \bibinfo {pages} {2581} (\bibinfo {year} {2012})}\BibitemShut
	{NoStop}%
	\bibitem [{\citenamefont {Braun}\ \emph {et~al.}(2014)\citenamefont {Braun},
		\citenamefont {Jian}, \citenamefont {Pinel},\ and\ \citenamefont
		{Treps}}]{braun-pra-2014}%
	\BibitemOpen
	\bibfield  {author} {\bibinfo {author} {\bibfnamefont {D.}~\bibnamefont
			{Braun}}, \bibinfo {author} {\bibfnamefont {P.}~\bibnamefont {Jian}},
		\bibinfo {author} {\bibfnamefont {O.}~\bibnamefont {Pinel}},\ and\ \bibinfo
		{author} {\bibfnamefont {N.}~\bibnamefont {Treps}},\ }\bibfield  {title}
	{\bibinfo {title} {Precision measurements with photon-subtracted or
			photon-added gaussian states},\ }\href
	{https://doi.org/10.1103/PhysRevA.90.013821} {\bibfield  {journal} {\bibinfo
			{journal} {Phys. Rev. A}\ }\textbf {\bibinfo {volume} {90}},\ \bibinfo
		{pages} {013821} (\bibinfo {year} {2014})}\BibitemShut {NoStop}%
	\bibitem [{\citenamefont {Ouyang}\ \emph {et~al.}(2016)\citenamefont {Ouyang},
		\citenamefont {Wang},\ and\ \citenamefont {Zhang}}]{josab-2016}%
	\BibitemOpen
	\bibfield  {author} {\bibinfo {author} {\bibfnamefont {Y.}~\bibnamefont
			{Ouyang}}, \bibinfo {author} {\bibfnamefont {S.}~\bibnamefont {Wang}},\ and\
		\bibinfo {author} {\bibfnamefont {L.}~\bibnamefont {Zhang}},\ }\bibfield
	{title} {\bibinfo {title} {Quantum optical interferometry via the
			photon-added two-mode squeezed vacuum states},\ }\href
	{https://doi.org/10.1364/JOSAB.33.001373} {\bibfield  {journal} {\bibinfo
			{journal} {J. Opt. Soc. Am. B}\ }\textbf {\bibinfo {volume} {33}},\ \bibinfo
		{pages} {1373} (\bibinfo {year} {2016})}\BibitemShut {NoStop}%
	\bibitem [{\citenamefont {Zhang}\ \emph {et~al.}(2021)\citenamefont {Zhang},
		\citenamefont {Ye}, \citenamefont {Wei}, \citenamefont {Xia}, \citenamefont
		{Chang}, \citenamefont {Liao},\ and\ \citenamefont
		{Hu}}]{pra-catalysis-2021}%
	\BibitemOpen
	\bibfield  {author} {\bibinfo {author} {\bibfnamefont {H.}~\bibnamefont
			{Zhang}}, \bibinfo {author} {\bibfnamefont {W.}~\bibnamefont {Ye}}, \bibinfo
		{author} {\bibfnamefont {C.}~\bibnamefont {Wei}}, \bibinfo {author}
		{\bibfnamefont {Y.}~\bibnamefont {Xia}}, \bibinfo {author} {\bibfnamefont
			{S.}~\bibnamefont {Chang}}, \bibinfo {author} {\bibfnamefont
			{Z.}~\bibnamefont {Liao}},\ and\ \bibinfo {author} {\bibfnamefont
			{L.}~\bibnamefont {Hu}},\ }\bibfield  {title} {\bibinfo {title} {Improved
			phase sensitivity in a quantum optical interferometer based on multiphoton
			catalytic two-mode squeezed vacuum states},\ }\href
	{https://doi.org/10.1103/PhysRevA.103.013705} {\bibfield  {journal} {\bibinfo
			{journal} {Phys. Rev. A}\ }\textbf {\bibinfo {volume} {103}},\ \bibinfo
		{pages} {013705} (\bibinfo {year} {2021})}\BibitemShut {NoStop}%
	\bibitem [{\citenamefont {Tan}\ \emph {et~al.}(2008)\citenamefont {Tan},
		\citenamefont {Erkmen}, \citenamefont {Giovannetti}, \citenamefont {Guha},
		\citenamefont {Lloyd}, \citenamefont {Maccone}, \citenamefont {Pirandola},\
		and\ \citenamefont {Shapiro}}]{ill2008}%
	\BibitemOpen
	\bibfield  {author} {\bibinfo {author} {\bibfnamefont {S.-H.}\ \bibnamefont
			{Tan}}, \bibinfo {author} {\bibfnamefont {B.~I.}\ \bibnamefont {Erkmen}},
		\bibinfo {author} {\bibfnamefont {V.}~\bibnamefont {Giovannetti}}, \bibinfo
		{author} {\bibfnamefont {S.}~\bibnamefont {Guha}}, \bibinfo {author}
		{\bibfnamefont {S.}~\bibnamefont {Lloyd}}, \bibinfo {author} {\bibfnamefont
			{L.}~\bibnamefont {Maccone}}, \bibinfo {author} {\bibfnamefont
			{S.}~\bibnamefont {Pirandola}},\ and\ \bibinfo {author} {\bibfnamefont
			{J.~H.}\ \bibnamefont {Shapiro}},\ }\bibfield  {title} {\bibinfo {title}
		{Quantum illumination with gaussian states},\ }\href
	{https://doi.org/10.1103/PhysRevLett.101.253601} {\bibfield  {journal}
		{\bibinfo  {journal} {Phys. Rev. Lett.}\ }\textbf {\bibinfo {volume} {101}},\
		\bibinfo {pages} {253601} (\bibinfo {year} {2008})}\BibitemShut {NoStop}%
	\bibitem [{\citenamefont {Lopaeva}\ \emph {et~al.}(2013)\citenamefont
		{Lopaeva}, \citenamefont {Ruo~Berchera}, \citenamefont {Degiovanni},
		\citenamefont {Olivares}, \citenamefont {Brida},\ and\ \citenamefont
		{Genovese}}]{ill2013}%
	\BibitemOpen
	\bibfield  {author} {\bibinfo {author} {\bibfnamefont {E.~D.}\ \bibnamefont
			{Lopaeva}}, \bibinfo {author} {\bibfnamefont {I.}~\bibnamefont
			{Ruo~Berchera}}, \bibinfo {author} {\bibfnamefont {I.~P.}\ \bibnamefont
			{Degiovanni}}, \bibinfo {author} {\bibfnamefont {S.}~\bibnamefont
			{Olivares}}, \bibinfo {author} {\bibfnamefont {G.}~\bibnamefont {Brida}},\
		and\ \bibinfo {author} {\bibfnamefont {M.}~\bibnamefont {Genovese}},\
	}\bibfield  {title} {\bibinfo {title} {Experimental realization of quantum
			illumination},\ }\href {https://doi.org/10.1103/PhysRevLett.110.153603}
	{\bibfield  {journal} {\bibinfo  {journal} {Phys. Rev. Lett.}\ }\textbf
		{\bibinfo {volume} {110}},\ \bibinfo {pages} {153603} (\bibinfo {year}
		{2013})}\BibitemShut {NoStop}%
	\bibitem [{\citenamefont {Kumar}\ \emph {et~al.}(2022)\citenamefont {Kumar},
		\citenamefont {Rishabh},\ and\ \citenamefont {Arora}}]{metro22}%
	\BibitemOpen
	\bibfield  {author} {\bibinfo {author} {\bibfnamefont {C.}~\bibnamefont
			{Kumar}}, \bibinfo {author} {\bibnamefont {Rishabh}},\ and\ \bibinfo {author}
		{\bibfnamefont {S.}~\bibnamefont {Arora}},\ }\bibfield  {title} {\bibinfo
		{title} {Realistic non-gaussian-operation scheme in parity-detection-based
			mach-zehnder quantum interferometry},\ }\href
	{https://doi.org/10.1103/PhysRevA.105.052437} {\bibfield  {journal} {\bibinfo
			{journal} {Phys. Rev. A}\ }\textbf {\bibinfo {volume} {105}},\ \bibinfo
		{pages} {052437} (\bibinfo {year} {2022})}\BibitemShut {NoStop}%
	\bibitem [{\citenamefont {Kumar}\ \emph
		{et~al.}(2023{\natexlab{a}})\citenamefont {Kumar}, \citenamefont {Rishabh},\
		and\ \citenamefont {Arora}}]{metro-thermal-arxiv}%
	\BibitemOpen
	\bibfield  {author} {\bibinfo {author} {\bibfnamefont {C.}~\bibnamefont
			{Kumar}}, \bibinfo {author} {\bibnamefont {Rishabh}},\ and\ \bibinfo {author}
		{\bibfnamefont {S.}~\bibnamefont {Arora}},\ }\bibfield  {title} {\bibinfo
		{title} {Enhanced phase estimation in parity-detection-based mach–zehnder
			interferometer using non-gaussian two-mode squeezed thermal input state},\
	}\href {https://doi.org/https://doi.org/10.1002/andp.202300117} {\bibfield
		{journal} {\bibinfo  {journal} {Annalen der Physik}\ }\textbf {\bibinfo
			{volume} {535}},\ \bibinfo {pages} {2300117} (\bibinfo {year}
		{2023}{\natexlab{a}})}\BibitemShut {NoStop}%
	\bibitem [{\citenamefont {Kumar}\ \emph
		{et~al.}(2023{\natexlab{b}})\citenamefont {Kumar}, \citenamefont {Rishabh},
		\citenamefont {Sharma},\ and\ \citenamefont {Arora}}]{ngsvs-arxiv}%
	\BibitemOpen
	\bibfield  {author} {\bibinfo {author} {\bibfnamefont {C.}~\bibnamefont
			{Kumar}}, \bibinfo {author} {\bibnamefont {Rishabh}}, \bibinfo {author}
		{\bibfnamefont {M.}~\bibnamefont {Sharma}},\ and\ \bibinfo {author}
		{\bibfnamefont {S.}~\bibnamefont {Arora}},\ }\bibfield  {title} {\bibinfo
		{title} {Parity-detection-based mach-zehnder interferometry with coherent and
			non-gaussian squeezed vacuum states as inputs},\ }\href
	{https://doi.org/10.1103/PhysRevA.108.012605} {\bibfield  {journal} {\bibinfo
			{journal} {Phys. Rev. A}\ }\textbf {\bibinfo {volume} {108}},\ \bibinfo
		{pages} {012605} (\bibinfo {year} {2023}{\natexlab{b}})}\BibitemShut
	{NoStop}%
	\bibitem [{\citenamefont {Huang}\ \emph {et~al.}(2013)\citenamefont {Huang},
		\citenamefont {He}, \citenamefont {Fang},\ and\ \citenamefont
		{Zeng}}]{qkd-pra-2013}%
	\BibitemOpen
	\bibfield  {author} {\bibinfo {author} {\bibfnamefont {P.}~\bibnamefont
			{Huang}}, \bibinfo {author} {\bibfnamefont {G.}~\bibnamefont {He}}, \bibinfo
		{author} {\bibfnamefont {J.}~\bibnamefont {Fang}},\ and\ \bibinfo {author}
		{\bibfnamefont {G.}~\bibnamefont {Zeng}},\ }\bibfield  {title} {\bibinfo
		{title} {Performance improvement of continuous-variable quantum key
			distribution via photon subtraction},\ }\href
	{https://doi.org/10.1103/PhysRevA.87.012317} {\bibfield  {journal} {\bibinfo
			{journal} {Phys. Rev. A}\ }\textbf {\bibinfo {volume} {87}},\ \bibinfo
		{pages} {012317} (\bibinfo {year} {2013})}\BibitemShut {NoStop}%
	\bibitem [{\citenamefont {Guo}\ \emph {et~al.}(2019)\citenamefont {Guo},
		\citenamefont {Ye}, \citenamefont {Zhong},\ and\ \citenamefont
		{Liao}}]{qkd-pra-2019}%
	\BibitemOpen
	\bibfield  {author} {\bibinfo {author} {\bibfnamefont {Y.}~\bibnamefont
			{Guo}}, \bibinfo {author} {\bibfnamefont {W.}~\bibnamefont {Ye}}, \bibinfo
		{author} {\bibfnamefont {H.}~\bibnamefont {Zhong}},\ and\ \bibinfo {author}
		{\bibfnamefont {Q.}~\bibnamefont {Liao}},\ }\bibfield  {title} {\bibinfo
		{title} {Continuous-variable quantum key distribution with non-gaussian
			quantum catalysis},\ }\href {https://doi.org/10.1103/PhysRevA.99.032327}
	{\bibfield  {journal} {\bibinfo  {journal} {Phys. Rev. A}\ }\textbf {\bibinfo
			{volume} {99}},\ \bibinfo {pages} {032327} (\bibinfo {year}
		{2019})}\BibitemShut {NoStop}%
	\bibitem [{\citenamefont {Ye}\ \emph {et~al.}(2019)\citenamefont {Ye},
		\citenamefont {Zhong}, \citenamefont {Liao}, \citenamefont {Huang},
		\citenamefont {Hu},\ and\ \citenamefont {Guo}}]{qk2019}%
	\BibitemOpen
	\bibfield  {author} {\bibinfo {author} {\bibfnamefont {W.}~\bibnamefont
			{Ye}}, \bibinfo {author} {\bibfnamefont {H.}~\bibnamefont {Zhong}}, \bibinfo
		{author} {\bibfnamefont {Q.}~\bibnamefont {Liao}}, \bibinfo {author}
		{\bibfnamefont {D.}~\bibnamefont {Huang}}, \bibinfo {author} {\bibfnamefont
			{L.}~\bibnamefont {Hu}},\ and\ \bibinfo {author} {\bibfnamefont
			{Y.}~\bibnamefont {Guo}},\ }\bibfield  {title} {\bibinfo {title} {Improvement
			of self-referenced continuous-variable quantum key distribution with quantum
			photon catalysis},\ }\href {https://doi.org/10.1364/OE.27.017186} {\bibfield
		{journal} {\bibinfo  {journal} {Opt. Express}\ }\textbf {\bibinfo {volume}
			{27}},\ \bibinfo {pages} {17186} (\bibinfo {year} {2019})}\BibitemShut
	{NoStop}%
	\bibitem [{\citenamefont {Hu}\ \emph {et~al.}(2020)\citenamefont {Hu},
		\citenamefont {Al-amri}, \citenamefont {Liao},\ and\ \citenamefont
		{Zubairy}}]{zubairy-pra-2020}%
	\BibitemOpen
	\bibfield  {author} {\bibinfo {author} {\bibfnamefont {L.}~\bibnamefont
			{Hu}}, \bibinfo {author} {\bibfnamefont {M.}~\bibnamefont {Al-amri}},
		\bibinfo {author} {\bibfnamefont {Z.}~\bibnamefont {Liao}},\ and\ \bibinfo
		{author} {\bibfnamefont {M.~S.}\ \bibnamefont {Zubairy}},\ }\bibfield
	{title} {\bibinfo {title} {Continuous-variable quantum key distribution with
			non-gaussian operations},\ }\href
	{https://doi.org/10.1103/PhysRevA.102.012608} {\bibfield  {journal} {\bibinfo
			{journal} {Phys. Rev. A}\ }\textbf {\bibinfo {volume} {102}},\ \bibinfo
		{pages} {012608} (\bibinfo {year} {2020})}\BibitemShut {NoStop}%
	\bibitem [{\citenamefont {Ma}\ \emph {et~al.}(2018)\citenamefont {Ma},
		\citenamefont {Huang}, \citenamefont {Bai}, \citenamefont {Wang},
		\citenamefont {Bao},\ and\ \citenamefont {Zeng}}]{Ma-pra-2018}%
	\BibitemOpen
	\bibfield  {author} {\bibinfo {author} {\bibfnamefont {H.-X.}\ \bibnamefont
			{Ma}}, \bibinfo {author} {\bibfnamefont {P.}~\bibnamefont {Huang}}, \bibinfo
		{author} {\bibfnamefont {D.-Y.}\ \bibnamefont {Bai}}, \bibinfo {author}
		{\bibfnamefont {S.-Y.}\ \bibnamefont {Wang}}, \bibinfo {author}
		{\bibfnamefont {W.-S.}\ \bibnamefont {Bao}},\ and\ \bibinfo {author}
		{\bibfnamefont {G.-H.}\ \bibnamefont {Zeng}},\ }\bibfield  {title} {\bibinfo
		{title} {Continuous-variable measurement-device-independent quantum key
			distribution with photon subtraction},\ }\href
	{https://doi.org/10.1103/PhysRevA.97.042329} {\bibfield  {journal} {\bibinfo
			{journal} {Phys. Rev. A}\ }\textbf {\bibinfo {volume} {97}},\ \bibinfo
		{pages} {042329} (\bibinfo {year} {2018})}\BibitemShut {NoStop}%
	\bibitem [{\citenamefont {Kumar}\ \emph {et~al.}(2019)\citenamefont {Kumar},
		\citenamefont {Singh}, \citenamefont {Bose},\ and\ \citenamefont
		{Arvind}}]{chandan-pra-2019}%
	\BibitemOpen
	\bibfield  {author} {\bibinfo {author} {\bibfnamefont {C.}~\bibnamefont
			{Kumar}}, \bibinfo {author} {\bibfnamefont {J.}~\bibnamefont {Singh}},
		\bibinfo {author} {\bibfnamefont {S.}~\bibnamefont {Bose}},\ and\ \bibinfo
		{author} {\bibnamefont {Arvind}},\ }\bibfield  {title} {\bibinfo {title}
		{Coherence-assisted non-gaussian measurement-device-independent quantum key
			distribution},\ }\href {https://doi.org/10.1103/PhysRevA.100.052329}
	{\bibfield  {journal} {\bibinfo  {journal} {Phys. Rev. A}\ }\textbf {\bibinfo
			{volume} {100}},\ \bibinfo {pages} {052329} (\bibinfo {year}
		{2019})}\BibitemShut {NoStop}%
	\bibitem [{\citenamefont {Ye}\ \emph {et~al.}(2020)\citenamefont {Ye},
		\citenamefont {Zhong}, \citenamefont {Wu}, \citenamefont {Hu},\ and\
		\citenamefont {Guo}}]{catcvmdiqkd2020}%
	\BibitemOpen
	\bibfield  {author} {\bibinfo {author} {\bibfnamefont {W.}~\bibnamefont
			{Ye}}, \bibinfo {author} {\bibfnamefont {H.}~\bibnamefont {Zhong}}, \bibinfo
		{author} {\bibfnamefont {X.}~\bibnamefont {Wu}}, \bibinfo {author}
		{\bibfnamefont {L.}~\bibnamefont {Hu}},\ and\ \bibinfo {author}
		{\bibfnamefont {Y.}~\bibnamefont {Guo}},\ }\bibfield  {title} {\bibinfo
		{title} {Continuous-variable measurement-device-independent quantum key
			distribution via quantum catalysis},\ }\href
	{https://doi.org/10.1007/s11128-020-02859-3} {\bibfield  {journal} {\bibinfo
			{journal} {Quantum Information Processing}\ }\textbf {\bibinfo {volume}
			{19}},\ \bibinfo {pages} {346} (\bibinfo {year} {2020})}\BibitemShut
	{NoStop}%
	\bibitem [{\citenamefont {Singh}\ and\ \citenamefont
		{Bose}(2021)}]{catalysis2021}%
	\BibitemOpen
	\bibfield  {author} {\bibinfo {author} {\bibfnamefont {J.}~\bibnamefont
			{Singh}}\ and\ \bibinfo {author} {\bibfnamefont {S.}~\bibnamefont {Bose}},\
	}\bibfield  {title} {\bibinfo {title} {Non-gaussian operations in
			measurement-device-independent quantum key distribution},\ }\href
	{https://doi.org/10.1103/PhysRevA.104.052605} {\bibfield  {journal} {\bibinfo
			{journal} {Phys. Rev. A}\ }\textbf {\bibinfo {volume} {104}},\ \bibinfo
		{pages} {052605} (\bibinfo {year} {2021})}\BibitemShut {NoStop}%
	\bibitem [{\citenamefont {Kumar}\ \emph
		{et~al.}(2024{\natexlab{b}})\citenamefont {Kumar}, \citenamefont
		{Chatterjee},\ and\ \citenamefont {Arvind}}]{long-qkd}%
	\BibitemOpen
	\bibfield  {author} {\bibinfo {author} {\bibfnamefont {C.}~\bibnamefont
			{Kumar}}, \bibinfo {author} {\bibfnamefont {S.}~\bibnamefont {Chatterjee}},\
		and\ \bibinfo {author} {\bibnamefont {Arvind}},\ }\bibfield  {title}
	{\bibinfo {title} {Optimization of state parameters in displacement assisted
			photon subtracted measurement-device-independent quantum key distribution},\
	}\href {https://doi.org/10.48550/arXiv.2406.04270} {\bibfield  {journal}
		{\bibinfo  {journal} {arxiv.2406.04270}\ } (\bibinfo {year}
		{2024}{\natexlab{b}})}\BibitemShut {NoStop}%
	\bibitem [{\citenamefont {Zhong}\ \emph {et~al.}(2020)\citenamefont {Zhong},
		\citenamefont {Guo}, \citenamefont {Mao}, \citenamefont {Ye},\ and\
		\citenamefont {Huang}}]{virtualpc}%
	\BibitemOpen
	\bibfield  {author} {\bibinfo {author} {\bibfnamefont {H.}~\bibnamefont
			{Zhong}}, \bibinfo {author} {\bibfnamefont {Y.}~\bibnamefont {Guo}}, \bibinfo
		{author} {\bibfnamefont {Y.}~\bibnamefont {Mao}}, \bibinfo {author}
		{\bibfnamefont {W.}~\bibnamefont {Ye}},\ and\ \bibinfo {author}
		{\bibfnamefont {D.}~\bibnamefont {Huang}},\ }\bibfield  {title} {\bibinfo
		{title} {Virtual zero-photon catalysis for improving continuous-variable
			quantum key distribution via gaussian post-selection},\ }\href
	{https://doi.org/10.1038/s41598-020-73379-4} {\bibfield  {journal} {\bibinfo
			{journal} {Scientific Reports}\ }\textbf {\bibinfo {volume} {10}},\ \bibinfo
		{pages} {17526} (\bibinfo {year} {2020})}\BibitemShut {NoStop}%
	\bibitem [{\citenamefont {Wang}\ \emph {et~al.}(2020)\citenamefont {Wang},
		\citenamefont {Zou}, \citenamefont {Mao},\ and\ \citenamefont
		{Guo}}]{waterpc}%
	\BibitemOpen
	\bibfield  {author} {\bibinfo {author} {\bibfnamefont {Y.}~\bibnamefont
			{Wang}}, \bibinfo {author} {\bibfnamefont {S.}~\bibnamefont {Zou}}, \bibinfo
		{author} {\bibfnamefont {Y.}~\bibnamefont {Mao}},\ and\ \bibinfo {author}
		{\bibfnamefont {Y.}~\bibnamefont {Guo}},\ }\bibfield  {title} {\bibinfo
		{title} {Improving underwater continuous-variable
			measurement-device-independent quantum key distribution via zero-photon
			catalysis},\ }\bibfield  {journal} {\bibinfo  {journal} {Entropy}\ }\textbf
	{\bibinfo {volume} {22}},\ \href {https://doi.org/10.3390/e22050571}
	{10.3390/e22050571} (\bibinfo {year} {2020})\BibitemShut {NoStop}%
	\bibitem [{\citenamefont {Weedbrook}\ \emph {et~al.}(2012)\citenamefont
		{Weedbrook}, \citenamefont {Pirandola}, \citenamefont {Garc\'{\i}a-Patr\'on},
		\citenamefont {Cerf}, \citenamefont {Ralph}, \citenamefont {Shapiro},\ and\
		\citenamefont {Lloyd}}]{Weedbrook-rmp-2012}%
	\BibitemOpen
	\bibfield  {author} {\bibinfo {author} {\bibfnamefont {C.}~\bibnamefont
			{Weedbrook}}, \bibinfo {author} {\bibfnamefont {S.}~\bibnamefont
			{Pirandola}}, \bibinfo {author} {\bibfnamefont {R.}~\bibnamefont
			{Garc\'{\i}a-Patr\'on}}, \bibinfo {author} {\bibfnamefont {N.~J.}\
			\bibnamefont {Cerf}}, \bibinfo {author} {\bibfnamefont {T.~C.}\ \bibnamefont
			{Ralph}}, \bibinfo {author} {\bibfnamefont {J.~H.}\ \bibnamefont {Shapiro}},\
		and\ \bibinfo {author} {\bibfnamefont {S.}~\bibnamefont {Lloyd}},\ }\bibfield
	{title} {\bibinfo {title} {Gaussian quantum information},\ }\href
	{https://doi.org/10.1103/RevModPhys.84.621} {\bibfield  {journal} {\bibinfo
			{journal} {Rev. Mod. Phys.}\ }\textbf {\bibinfo {volume} {84}},\ \bibinfo
		{pages} {621} (\bibinfo {year} {2012})}\BibitemShut {NoStop}%
	\bibitem [{\citenamefont {Olivares}(2012)}]{olivares-2012}%
	\BibitemOpen
	\bibfield  {author} {\bibinfo {author} {\bibfnamefont {S.}~\bibnamefont
			{Olivares}},\ }\bibfield  {title} {\bibinfo {title} {Quantum optics in the
			phase space},\ }\href {https://doi.org/10.1140/epjst/e2012-01532-4}
	{\bibfield  {journal} {\bibinfo  {journal} {The European Physical Journal
				Special Topics}\ }\textbf {\bibinfo {volume} {203}},\ \bibinfo {pages} {3}
		(\bibinfo {year} {2012})}\BibitemShut {NoStop}%
	\bibitem [{\citenamefont {Garc\'{\i}a-Patr\'on}\ and\ \citenamefont
		{Cerf}(2006)}]{Cerf-prl-2006}%
	\BibitemOpen
	\bibfield  {author} {\bibinfo {author} {\bibfnamefont {R.}~\bibnamefont
			{Garc\'{\i}a-Patr\'on}}\ and\ \bibinfo {author} {\bibfnamefont {N.~J.}\
			\bibnamefont {Cerf}},\ }\bibfield  {title} {\bibinfo {title} {Unconditional
			optimality of gaussian attacks against continuous-variable quantum key
			distribution},\ }\href {https://doi.org/10.1103/PhysRevLett.97.190503}
	{\bibfield  {journal} {\bibinfo  {journal} {Phys. Rev. Lett.}\ }\textbf
		{\bibinfo {volume} {97}},\ \bibinfo {pages} {190503} (\bibinfo {year}
		{2006})}\BibitemShut {NoStop}%
	\bibitem [{\citenamefont {Ottaviani}\ \emph {et~al.}(2015)\citenamefont
		{Ottaviani}, \citenamefont {Spedalieri}, \citenamefont {Braunstein},\ and\
		\citenamefont {Pirandola}}]{ottaviani-pra-2015}%
	\BibitemOpen
	\bibfield  {author} {\bibinfo {author} {\bibfnamefont {C.}~\bibnamefont
			{Ottaviani}}, \bibinfo {author} {\bibfnamefont {G.}~\bibnamefont
			{Spedalieri}}, \bibinfo {author} {\bibfnamefont {S.~L.}\ \bibnamefont
			{Braunstein}},\ and\ \bibinfo {author} {\bibfnamefont {S.}~\bibnamefont
			{Pirandola}},\ }\bibfield  {title} {\bibinfo {title} {Continuous-variable
			quantum cryptography with an untrusted relay: Detailed security analysis of
			the symmetric configuration},\ }\href
	{https://doi.org/10.1103/PhysRevA.91.022320} {\bibfield  {journal} {\bibinfo
			{journal} {Phys. Rev. A}\ }\textbf {\bibinfo {volume} {91}},\ \bibinfo
		{pages} {022320} (\bibinfo {year} {2015})}\BibitemShut {NoStop}%
	\bibitem [{\citenamefont {Grasselli}(2021)}]{Grasselli2021}%
	\BibitemOpen
	\bibfield  {author} {\bibinfo {author} {\bibfnamefont {F.}~\bibnamefont
			{Grasselli}},\ }\bibinfo {title} {Beyond point-to-point quantum key
		distribution},\ in\ \href {https://doi.org/10.1007/978-3-030-64360-7_6}
	{\emph {\bibinfo {booktitle} {Quantum Cryptography: From Key Distribution to
				Conference Key Agreement}}}\ (\bibinfo  {publisher} {Springer International
		Publishing},\ \bibinfo {address} {Cham},\ \bibinfo {year} {2021})\ pp.\
	\bibinfo {pages} {83--104}\BibitemShut {NoStop}%
	\bibitem [{\citenamefont {Liao}\ \emph {et~al.}(2017)\citenamefont {Liao},
		\citenamefont {Cai}, \citenamefont {Liu}, \citenamefont {Zhang},
		\citenamefont {Li}, \citenamefont {Ren}, \citenamefont {Yin}, \citenamefont
		{Shen}, \citenamefont {Cao}, \citenamefont {Li}, \citenamefont {Li},
		\citenamefont {Chen}, \citenamefont {Sun}, \citenamefont {Jia}, \citenamefont
		{Wu}, \citenamefont {Jiang}, \citenamefont {Wang}, \citenamefont {Huang},
		\citenamefont {Wang}, \citenamefont {Zhou}, \citenamefont {Deng},
		\citenamefont {Xi}, \citenamefont {Ma}, \citenamefont {Hu}, \citenamefont
		{Zhang}, \citenamefont {Chen}, \citenamefont {Liu}, \citenamefont {Wang},
		\citenamefont {Zhu}, \citenamefont {Lu}, \citenamefont {Shu}, \citenamefont
		{Peng}, \citenamefont {Wang},\ and\ \citenamefont {Pan}}]{Liao2017}%
	\BibitemOpen
	\bibfield  {author} {\bibinfo {author} {\bibfnamefont {S.-K.}\ \bibnamefont
			{Liao}}, \bibinfo {author} {\bibfnamefont {W.-Q.}\ \bibnamefont {Cai}},
		\bibinfo {author} {\bibfnamefont {W.-Y.}\ \bibnamefont {Liu}}, \bibinfo
		{author} {\bibfnamefont {L.}~\bibnamefont {Zhang}}, \bibinfo {author}
		{\bibfnamefont {Y.}~\bibnamefont {Li}}, \bibinfo {author} {\bibfnamefont
			{J.-G.}\ \bibnamefont {Ren}}, \bibinfo {author} {\bibfnamefont
			{J.}~\bibnamefont {Yin}}, \bibinfo {author} {\bibfnamefont {Q.}~\bibnamefont
			{Shen}}, \bibinfo {author} {\bibfnamefont {Y.}~\bibnamefont {Cao}}, \bibinfo
		{author} {\bibfnamefont {Z.-P.}\ \bibnamefont {Li}}, \bibinfo {author}
		{\bibfnamefont {F.-Z.}\ \bibnamefont {Li}}, \bibinfo {author} {\bibfnamefont
			{X.-W.}\ \bibnamefont {Chen}}, \bibinfo {author} {\bibfnamefont {L.-H.}\
			\bibnamefont {Sun}}, \bibinfo {author} {\bibfnamefont {J.-J.}\ \bibnamefont
			{Jia}}, \bibinfo {author} {\bibfnamefont {J.-C.}\ \bibnamefont {Wu}},
		\bibinfo {author} {\bibfnamefont {X.-J.}\ \bibnamefont {Jiang}}, \bibinfo
		{author} {\bibfnamefont {J.-F.}\ \bibnamefont {Wang}}, \bibinfo {author}
		{\bibfnamefont {Y.-M.}\ \bibnamefont {Huang}}, \bibinfo {author}
		{\bibfnamefont {Q.}~\bibnamefont {Wang}}, \bibinfo {author} {\bibfnamefont
			{Y.-L.}\ \bibnamefont {Zhou}}, \bibinfo {author} {\bibfnamefont
			{L.}~\bibnamefont {Deng}}, \bibinfo {author} {\bibfnamefont {T.}~\bibnamefont
			{Xi}}, \bibinfo {author} {\bibfnamefont {L.}~\bibnamefont {Ma}}, \bibinfo
		{author} {\bibfnamefont {T.}~\bibnamefont {Hu}}, \bibinfo {author}
		{\bibfnamefont {Q.}~\bibnamefont {Zhang}}, \bibinfo {author} {\bibfnamefont
			{Y.-A.}\ \bibnamefont {Chen}}, \bibinfo {author} {\bibfnamefont {N.-L.}\
			\bibnamefont {Liu}}, \bibinfo {author} {\bibfnamefont {X.-B.}\ \bibnamefont
			{Wang}}, \bibinfo {author} {\bibfnamefont {Z.-C.}\ \bibnamefont {Zhu}},
		\bibinfo {author} {\bibfnamefont {C.-Y.}\ \bibnamefont {Lu}}, \bibinfo
		{author} {\bibfnamefont {R.}~\bibnamefont {Shu}}, \bibinfo {author}
		{\bibfnamefont {C.-Z.}\ \bibnamefont {Peng}}, \bibinfo {author}
		{\bibfnamefont {J.-Y.}\ \bibnamefont {Wang}},\ and\ \bibinfo {author}
		{\bibfnamefont {J.-W.}\ \bibnamefont {Pan}},\ }\bibfield  {title} {\bibinfo
		{title} {Satellite-to-ground quantum key distribution},\ }\href
	{https://doi.org/10.1038/nature23655} {\bibfield  {journal} {\bibinfo
			{journal} {Nature}\ }\textbf {\bibinfo {volume} {549}},\ \bibinfo {pages}
		{43} (\bibinfo {year} {2017})}\BibitemShut {NoStop}%
	\bibitem [{\citenamefont {Grosshans}\ \emph {et~al.}(2003)\citenamefont
		{Grosshans}, \citenamefont {Cerf}, \citenamefont {Wenger}, \citenamefont
		{Tualle-Brouri},\ and\ \citenamefont {Grangier}}]{oneway}%
	\BibitemOpen
	\bibfield  {author} {\bibinfo {author} {\bibfnamefont {F.}~\bibnamefont
			{Grosshans}}, \bibinfo {author} {\bibfnamefont {N.~J.}\ \bibnamefont {Cerf}},
		\bibinfo {author} {\bibfnamefont {J.}~\bibnamefont {Wenger}}, \bibinfo
		{author} {\bibfnamefont {R.}~\bibnamefont {Tualle-Brouri}},\ and\ \bibinfo
		{author} {\bibfnamefont {P.}~\bibnamefont {Grangier}},\ }\bibfield  {title}
	{\bibinfo {title} {Virtual entanglement and reconciliation protocols for
			quantum cryptography with continuous variables},\ }\href@noop {} {\bibfield
		{journal} {\bibinfo  {journal} {Quantum Info. Comput.}\ }\textbf {\bibinfo
			{volume} {3}},\ \bibinfo {pages} {535–552} (\bibinfo {year}
		{2003})}\BibitemShut {NoStop}%
	\bibitem [{\citenamefont {Weedbrook}\ \emph {et~al.}(2004)\citenamefont
		{Weedbrook}, \citenamefont {Lance}, \citenamefont {Bowen}, \citenamefont
		{Symul}, \citenamefont {Ralph},\ and\ \citenamefont {Lam}}]{prl2004}%
	\BibitemOpen
	\bibfield  {author} {\bibinfo {author} {\bibfnamefont {C.}~\bibnamefont
			{Weedbrook}}, \bibinfo {author} {\bibfnamefont {A.~M.}\ \bibnamefont
			{Lance}}, \bibinfo {author} {\bibfnamefont {W.~P.}\ \bibnamefont {Bowen}},
		\bibinfo {author} {\bibfnamefont {T.}~\bibnamefont {Symul}}, \bibinfo
		{author} {\bibfnamefont {T.~C.}\ \bibnamefont {Ralph}},\ and\ \bibinfo
		{author} {\bibfnamefont {P.~K.}\ \bibnamefont {Lam}},\ }\bibfield  {title}
	{\bibinfo {title} {Quantum cryptography without switching},\ }\href
	{https://doi.org/10.1103/PhysRevLett.93.170504} {\bibfield  {journal}
		{\bibinfo  {journal} {Phys. Rev. Lett.}\ }\textbf {\bibinfo {volume} {93}},\
		\bibinfo {pages} {170504} (\bibinfo {year} {2004})}\BibitemShut {NoStop}%
	\bibitem [{\citenamefont {Devetak}\ and\ \citenamefont
		{Winter}(2005)}]{winter}%
	\BibitemOpen
	\bibfield  {author} {\bibinfo {author} {\bibfnamefont {I.}~\bibnamefont
			{Devetak}}\ and\ \bibinfo {author} {\bibfnamefont {A.}~\bibnamefont
			{Winter}},\ }\bibfield  {title} {\bibinfo {title} {Distillation of secret key
			and entanglement from quantum states},\ }\href
	{https://doi.org/10.1098/rspa.2004.1372} {\bibfield  {journal} {\bibinfo
			{journal} {Proceedings of the Royal Society A: Mathematical, Physical and
				Engineering Sciences}\ }\textbf {\bibinfo {volume} {461}},\ \bibinfo {pages}
		{207} (\bibinfo {year} {2005})}\BibitemShut {NoStop}%
	\bibitem [{\citenamefont {Genoni}\ \emph {et~al.}(2008)\citenamefont {Genoni},
		\citenamefont {Paris},\ and\ \citenamefont {Banaszek}}]{non-G}%
	\BibitemOpen
	\bibfield  {author} {\bibinfo {author} {\bibfnamefont {M.~G.}\ \bibnamefont
			{Genoni}}, \bibinfo {author} {\bibfnamefont {M.~G.~A.}\ \bibnamefont
			{Paris}},\ and\ \bibinfo {author} {\bibfnamefont {K.}~\bibnamefont
			{Banaszek}},\ }\bibfield  {title} {\bibinfo {title} {Quantifying the
			non-gaussian character of a quantum state by quantum relative entropy},\
	}\href {https://doi.org/10.1103/PhysRevA.78.060303} {\bibfield  {journal}
		{\bibinfo  {journal} {Phys. Rev. A}\ }\textbf {\bibinfo {volume} {78}},\
		\bibinfo {pages} {060303} (\bibinfo {year} {2008})}\BibitemShut {NoStop}%
	\bibitem [{\citenamefont {Arvind}\ and\ \citenamefont
		{Mukunda}(1999)}]{am2002}%
	\BibitemOpen
	\bibfield  {author} {\bibinfo {author} {\bibnamefont {Arvind}}\ and\ \bibinfo
		{author} {\bibfnamefont {N.}~\bibnamefont {Mukunda}},\ }\bibfield  {title}
	{\bibinfo {title} {Bell's inequalities, multiphoton states and phase space
			distributions},\ }\href
	{https://doi.org/https://doi.org/10.1016/S0375-9601(99)00471-5} {\bibfield
		{journal} {\bibinfo  {journal} {Physics Letters A}\ }\textbf {\bibinfo
			{volume} {259}},\ \bibinfo {pages} {421} (\bibinfo {year}
		{1999})}\BibitemShut {NoStop}%
	\bibitem [{\citenamefont {Kumar}\ \emph {et~al.}(2021)\citenamefont {Kumar},
		\citenamefont {Saxena},\ and\ \citenamefont {Arvind}}]{nonlocality}%
	\BibitemOpen
	\bibfield  {author} {\bibinfo {author} {\bibfnamefont {C.}~\bibnamefont
			{Kumar}}, \bibinfo {author} {\bibfnamefont {G.}~\bibnamefont {Saxena}},\ and\
		\bibinfo {author} {\bibnamefont {Arvind}},\ }\bibfield  {title} {\bibinfo
		{title} {Continuous-variable clauser-horne bell-type inequality: A tool to
			unearth the nonlocality of continuous-variable quantum-optical systems},\
	}\href {https://doi.org/10.1103/PhysRevA.103.042224} {\bibfield  {journal}
		{\bibinfo  {journal} {Phys. Rev. A}\ }\textbf {\bibinfo {volume} {103}},\
		\bibinfo {pages} {042224} (\bibinfo {year} {2021})}\BibitemShut {NoStop}%
	\bibitem [{\citenamefont {Hillery}(1989)}]{Hillery}%
	\BibitemOpen
	\bibfield  {author} {\bibinfo {author} {\bibfnamefont {M.}~\bibnamefont
			{Hillery}},\ }\bibfield  {title} {\bibinfo {title} {Sum and difference
			squeezing of the electromagnetic field},\ }\href
	{https://doi.org/10.1103/PhysRevA.40.3147} {\bibfield  {journal} {\bibinfo
			{journal} {Phys. Rev. A}\ }\textbf {\bibinfo {volume} {40}},\ \bibinfo
		{pages} {3147} (\bibinfo {year} {1989})}\BibitemShut {NoStop}%
	\bibitem [{\citenamefont {Mandel}(1979)}]{Mandel:79}%
	\BibitemOpen
	\bibfield  {author} {\bibinfo {author} {\bibfnamefont {L.}~\bibnamefont
			{Mandel}},\ }\bibfield  {title} {\bibinfo {title} {Sub-poissonian photon
			statistics in resonance fluorescence},\ }\href
	{https://doi.org/10.1364/OL.4.000205} {\bibfield  {journal} {\bibinfo
			{journal} {Opt. Lett.}\ }\textbf {\bibinfo {volume} {4}},\ \bibinfo {pages}
		{205} (\bibinfo {year} {1979})}\BibitemShut {NoStop}%
	\bibitem [{\citenamefont {Lee}(1990)}]{antibunching}%
	\BibitemOpen
	\bibfield  {author} {\bibinfo {author} {\bibfnamefont {C.~T.}\ \bibnamefont
			{Lee}},\ }\bibfield  {title} {\bibinfo {title} {Many-photon antibunching in
			generalized pair coherent states},\ }\href
	{https://doi.org/10.1103/PhysRevA.41.1569} {\bibfield  {journal} {\bibinfo
			{journal} {Phys. Rev. A}\ }\textbf {\bibinfo {volume} {41}},\ \bibinfo
		{pages} {1569} (\bibinfo {year} {1990})}\BibitemShut {NoStop}%
	\bibitem [{\citenamefont {Jing}\ \emph {et~al.}(2021)\citenamefont {Jing},
		\citenamefont {Liu}, \citenamefont {Kong},\ and\ \citenamefont
		{He}}]{nlamdi}%
	\BibitemOpen
	\bibfield  {author} {\bibinfo {author} {\bibfnamefont {F.}~\bibnamefont
			{Jing}}, \bibinfo {author} {\bibfnamefont {W.}~\bibnamefont {Liu}}, \bibinfo
		{author} {\bibfnamefont {L.}~\bibnamefont {Kong}},\ and\ \bibinfo {author}
		{\bibfnamefont {C.}~\bibnamefont {He}},\ }\bibfield  {title} {\bibinfo
		{title} {Improving the performance of continuous-variable
			measurement-device-independent quantum key distribution via a noiseless
			linear amplifier},\ }\bibfield  {journal} {\bibinfo  {journal} {Entropy}\
	}\textbf {\bibinfo {volume} {23}},\ \href {https://doi.org/10.3390/e23121691}
	{10.3390/e23121691} (\bibinfo {year} {2021})\BibitemShut {NoStop}%
	\bibitem [{\citenamefont {Kong}\ \emph {et~al.}(2022)\citenamefont {Kong},
		\citenamefont {Liu}, \citenamefont {Jing}, \citenamefont {Zhang},
		\citenamefont {Qi},\ and\ \citenamefont {He}}]{s-mdi}%
	\BibitemOpen
	\bibfield  {author} {\bibinfo {author} {\bibfnamefont {L.}~\bibnamefont
			{Kong}}, \bibinfo {author} {\bibfnamefont {W.}~\bibnamefont {Liu}}, \bibinfo
		{author} {\bibfnamefont {F.}~\bibnamefont {Jing}}, \bibinfo {author}
		{\bibfnamefont {Z.-K.}\ \bibnamefont {Zhang}}, \bibinfo {author}
		{\bibfnamefont {J.}~\bibnamefont {Qi}},\ and\ \bibinfo {author}
		{\bibfnamefont {C.}~\bibnamefont {He}},\ }\bibfield  {title} {\bibinfo
		{title} {Improvement of a continuous-variable measurement-device-independent
			quantum key distribution system via quantum scissors},\ }\href
	{https://doi.org/10.1088/1674-1056/ac6dba} {\bibfield  {journal} {\bibinfo
			{journal} {Chinese Physics B}\ }\textbf {\bibinfo {volume} {31}},\ \bibinfo
		{pages} {090304} (\bibinfo {year} {2022})}\BibitemShut {NoStop}%
	\bibitem [{\citenamefont {Lee}\ and\ \citenamefont
		{Nha}(2013)}]{betterresource}%
	\BibitemOpen
	\bibfield  {author} {\bibinfo {author} {\bibfnamefont {J.}~\bibnamefont
			{Lee}}\ and\ \bibinfo {author} {\bibfnamefont {H.}~\bibnamefont {Nha}},\
	}\bibfield  {title} {\bibinfo {title} {Entanglement distillation for
			continuous variables in a thermal environment: Effectiveness of a
			non-gaussian operation},\ }\href {https://doi.org/10.1103/PhysRevA.87.032307}
	{\bibfield  {journal} {\bibinfo  {journal} {Phys. Rev. A}\ }\textbf {\bibinfo
			{volume} {87}},\ \bibinfo {pages} {032307} (\bibinfo {year}
		{2013})}\BibitemShut {NoStop}%
\end{thebibliography}
%

\end{document}